\documentclass[11pt,preprint]{aastex}

\shortauthors{Hoard et al.}
\shorttitle{New White Dwarf Dust Disks}

\begin{document}

\title{The WIRED Survey. IV. New Dust Disks from the\\McCook \& Sion White Dwarf Catalog}
 
\author{D.~W.\ Hoard\altaffilmark{1,2,3}, 
John H.\ Debes\altaffilmark{4},
Stefanie Wachter\altaffilmark{1},
David T.\ Leisawitz\altaffilmark{5},
and
Martin Cohen\altaffilmark{6}}

\altaffiltext{1}{Max Planck Institut f\"{u}r Astronomie, Heidelberg, 69117, Germany}
\altaffiltext{2}{Visiting Scientist, MPIA; email:\ {\tt hoard@mpia.de}}
\altaffiltext{3}{Eureka Scientific, Inc., Oakland, CA 94602, USA}
\altaffiltext{4}{Space Telescope Science Institute, Baltimore, MD 21218, USA}
\altaffiltext{5}{Goddard Space Flight Center, Greenbelt, MD 20771, USA}
\altaffiltext{6}{Monterey Institute for Research in Astronomy, Marina, CA 93933, USA}

\slugcomment{accepted for publication in The Astrophysical Journal, 26 Apr 2013}

% NOTE (5/3/2013): missing gms06 citation added during production

\begin{abstract}
We have compiled photometric data from the {\it Wide-field Infrared Survey Explorer} All Sky Survey and other archival sources for the more than 2200 objects in the original McCook \& Sion Catalog of Spectroscopically Identified White Dwarfs.
We applied color-selection criteria to identify 28 targets whose infrared spectral energy distributions depart from the expectation for the white dwarf photosphere alone.
Seven of these are previously known white dwarfs with circumstellar dust disks, five are known central stars of planetary nebulae, and six were excluded for being known binaries or having possible contamination of their infrared photometry.
We fit white dwarf models to the spectral energy distributions of the remaining ten targets, and find seven new candidates with infrared excess suggesting the presence of a circumstellar dust disk.
We compare the model dust disk properties for these new candidates with a comprehensive compilation of previously published parameters for known white dwarfs with dust disks.
It is possible that the current census of white dwarfs with dust disks that produce an excess detectable at K-band and shorter wavelengths 
is close to complete for the entire sample of known WDs to the detection limits of existing near-IR all-sky surveys.
The white dwarf dust disk candidates now being found using longer wavelength infrared data are drawn from a previously underrepresented region of parameter space, in which the dust disks are overall cooler, narrower in radial extent, and/or contain fewer emitting grains.
\end{abstract}

\keywords{circumstellar matter --- planetary systems --- surveys --- white dwarfs}

\section{Introduction}

Dust disks are common in a wide variety of astrophysical situations, 
including  
the central engines of quasars and active galactic nuclei (e.g., \citealt{rowan77,antonucci93}), 
the precursors of planetary system formation around protostars (e.g., \citealt{natta08})
and 
post-formation debris around young stars (e.g., \citealt{aumann85,ckk91}), 
and even 
the recently discovered largest ring of Saturn (e.g., \citealt{vsh09}). 
The first dust disk around a white dwarf (WD), G29-38, was discovered 
by virtue of its infrared (IR) excess over the WD photosphere \citep{zb87}, although it took another decade to cast aside all doubts that the excess was truly due to dust and not an unresolved brown dwarf companion \citep{kps97,kkb98}. 
\citet{ds02} and \citet{jura03} developed a model for the origin of WD dust disks involving tidal disruption of a comet or asteroid perturbed into the WD Roche lobe due to the gravitational influence of a remnant planetary system containing at least one massive planet \citep{dws12}. 
It wasn't until 2005 that the second WD with a dust disk, GD~362, was discovered \citep{bfj05,kvl05}. 
By the end of 2010, 20 dusty WDs were known (Table~5.1 in \citealt{farihi11}), largely owing to sensitive IR observations from the {\it Spitzer Space Telescope}.

Predating the discovery of dust around WDs, it was known that a small fraction of WDs show absorption lines of metals in their optical and UV photospheric spectra (e.g., \citealt{lwf83,sg83,zwk86}).  These lines originate from ``pollution'' of a WD's geometrically thin, high density (but non-degenerate) atmosphere, which in most objects is otherwise pure hydrogen or helium.  Gravitational settling times in hydrogen-rich (DA) WD atmospheres are very short (a few days to $\lesssim1000$~yr), so metals quickly diffuse out of the photosphere unless replenished. Thus, the observed metals were thought to be supplied by ongoing accretion from the ISM \citep{ska90}. This explanation was problematic; notably, the required accretion rate is high ($\gtrsim10^{8.5}$~g~s$^{-1}$ compared to $\sim10^{7}$~g~s$^{-1}$ expected from ISM accretion) and it is difficult to explain the relative elemental abundances of the accreted material, which do not match equilibrium ISM values (see \citealt{fbr10} and the review and discussion in Sections~5.2.4 and 5.6.6 of \citealt{farihi11}). 

It is a testament to the strength of the asteroid disruption model for WD dust disks that it also explains the metal-rich WDs, via accretion from circumstellar dust. 
\citet{zuckerman07} showed that the relative abundances of accreted metals in the dusty WD GD~362 closely match those of the terrestrial planets. 
Analyses of two metal polluted WDs (GD~61, NLTT~43806) suggest that the accreted dust derived from 
an asteroid whose origin was in 
the outer layers of a differentiated planet, in which the heaviest elements had sunk to the core, leaving a lithosphere rich in Ca and possibly water \citep{fbg11,zkd11}. 
Thus, observing WDs with dust disks is directly linked to determining how the chemical diversity of planetary systems can influence the probability that some planets support life.

In order to facilitate this, a large sample of WDs with dust disks is desired.
Consequently, we have been carrying out the {\it WISE} InfraRed Excesses around Degenerates (WIRED) Survey, utilizing photometry from the 
{\it Wide-field Infrared Survey Explorer} ({\it WISE\/}), a NASA medium class Explorer mission launched on 14 Dec 2009 \citep{wise10}.  
{\it WISE} mapped the entire sky at 3.4, 4.6, 12, and 22~$\mu$m (W1, W2, W3, and W4 bands, respectively) with $5\sigma$ point source sensitivities of approximately 0.08, 0.11, 1, and 6~mJy, respectively.  Complete sky coverage was achieved in mid-July 2010.  
The WIRED Survey has the goals of characterizing WD stars in the {\it WISE} bands, confirming objects known to have IR excess from past observations ({\it Spitzer}, 2MASS, UKIDSS, etc.), and revealing new examples of WDs with IR excess that can be attributed to unresolved companions or circumstellar debris disks.  We are utilizing target lists drawn from cataloged WD samples (e.g., from the Sloan Digital Sky Survey, McCook \& Sion, etc.).  To date, we have published results from WIRED for the SDSS WD sample that have nearly tripled the number of known WDs with circumstellar dust disks and increased the number of WD + brown dwarf binaries by almost an order of magnitude.  We now present initial results from examining the \citet{ms99} (henceforth, MS99) catalog of spectroscopically identified WDs, which resulted in the new identification of 
%REF-NUMBER 
seven
WDs with IR excess indicating the likely presence of a circumstellar dust disk.

\section{Targets and Data}

MS99 contains 2249 (optical) spectroscopically identified WDs.
The updated and online version of MS99, the Villanova University White Dwarf Catalog\footnote{See \url{http://www.astronomy.villanova.edu/WDCatalog/index.html}.} (henceforth, MSonline) currently contains over 14,000 entries.  For our purposes, we have used only those targets contained in the original print publication of MS99 (most of the new WDs listed in MSonline are objects discovered by SDSS and are covered in our previous WIRED paper; see \citealt{wired2}).
Since the publication of MS99, a number of the WDs were subsequently reclassified as non-WDs (e.g., quasars) or nonexistent (e.g., some WDs are listed twice in MS99 with different names, such as WD~2321-549 = WD~J2324-546), leaving 2202 viable targets. 
To further narrow the target list, we considered only the 1474 WDs from MS99 for which \citet{hoard07} found 2MASS near-IR detections.

As a first pass at identifying IR-bright (or otherwise ``interesting'') WDs from this input list, we utilized the following selection criteria for each target: 
(1) 2MASS-J, W1, and W2 photometry exists, 
(2) the {\it WISE} color index (W1$-$W2) $\geq$ $+0.3$~mag,
and 
(3) the signal-to-noise ratios of the W1 and W2 detections are both $\geq7$. 

\begin{figure*}[tb]
\epsscale{0.88}
\plotone{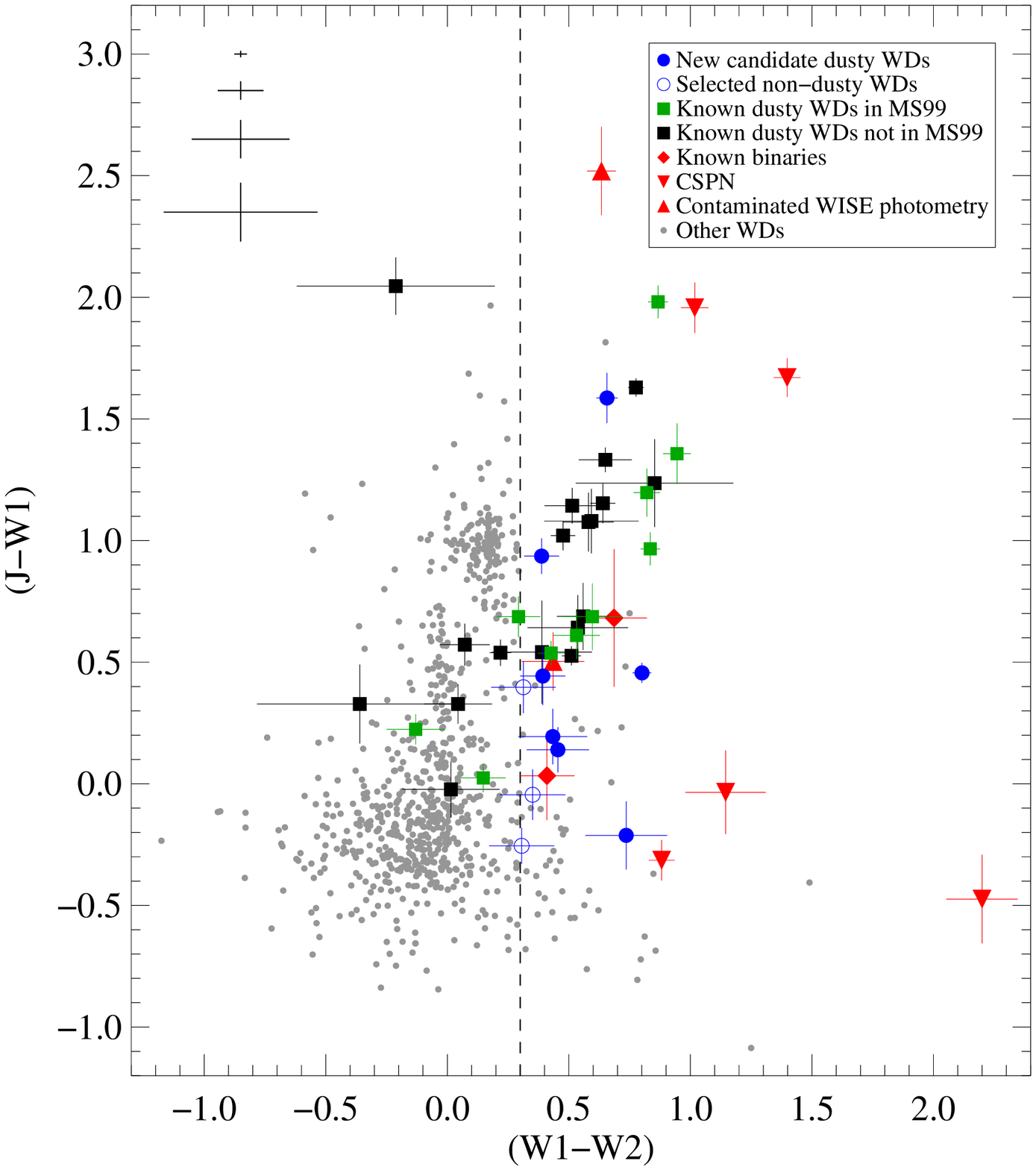}
\epsscale{1.0}
\figurenum{1}
\caption{%THIS IS THE COLOR VERSION OF THIS FIGURE FOR ONLINE ONLY.
Infrared color-color diagram of MS99 WDs with 2MASS J and {\it WISE} W1 and W2 detections.  
The vertical dashed line marks a (W1$-$W2) color of $+0.3$~mag, which was used in the target selection process.
Representative color index error bars are shown in the upper left; from top to bottom, 25\%, 50\%, 75\%, and 90\% of the detected targets have photometric uncertainties smaller than the indicated error bars.
The points are symbol-coded as follows:
new candidate WDs with dust disks (blue filled circles), 
selected targets that are not dust disk candidates (blue unfilled circles), 
known WDs with dust disks that are in (green squares) or not in (black squares) MS99, 
known unresolved binaries (red diamonds), 
central stars of planetary nebulae (red downward facing triangles), 
{\it WISE} photometry is contaminated (red upward facing triangles), 
and
remaining WDs that did not satisfy our selection criteria as targets-of-interest (small gray circles).
WD~1201+437 is also a known unresolved binary with (J$-$W1) $=$ $+3.8$~mag and, for clarity, is the only target not plotted here.
\label{f:ccdcolor}}
\end{figure*}

The IR color-color diagram of the MS99 WDs (see Figure~\ref{f:ccdcolor}) demonstrates that the 2nd criterion selects the majority of known dusty WDs, while avoiding the bulk of the ``uninteresting'' WDs.  The 3rd criterion excludes color-selected targets whose redness is spurious, resulting from low S/N photometry. 
A broad 10~$\mu$m silicate emission feature is a hallmark of circumstellar dust around WDs \citep{jfz09} and falls into the W3 band.  In principle, this could offer an additional selection criterion for identifying WDs with dust.  In practice, however, we found that due to the lower sensitivity of {\it WISE} in the W3 band compared to W1 and W2 (cf.\ while 67\% and 58\% of our input sample have a W1 and/or W2 detection, respectively, only 16\% have a W3 detection), almost 92\% of the targets that have a cataloged W3 detection are already selected by our 1st criterion.  
Only two of the targets with a W3 detection that were {\it not} selected by our 1st criterion have S/N $\geq7$ in W3, and both of them are unusable:\ WD~1919+145 is contaminated by a nearby source \citep{mkr07} and WD~2110+300 is in an unresolved binary with a G-type giant star ($\zeta$~Cygni; \citealt{gk92}).
In addition, closer inspection of the {\it WISE} images of our targets shows that as many as $\sim$50\% of the cataloged W3 ``detections'' (especially at low S/N) are probably unreliable (e.g., due to bright, structured background or nearby sources) and should be treated as upper limits (e.g., see Section~\ref{s:notes}).

Incidentally, there are some features of the IR color-color diagram that we will not discuss in detail, but are worth noting:

\begin{itemize}
\item{There are two principal loci of WDs (plotted as small gray circles):\ a large one centered around (J$-$W1)~$\approx-0.25$~mag, (W1$-$W2)~$\approx0$~mag, and a smaller one at (J$-$W1)~$\approx+1.0$~mag, (W1$-$W2)~$\approx+0.2$~mag.  The former is the locus of ``naked'' WDs, while the latter is the locus of unresolved binaries containing a WD and a low mass main sequence star.  We found a similar distribution of the majority of points in the r, i, J, W1, W2 color-color planes for the targets selected from the SDSS Data Release 7 preliminary WD catalog \citep{wired2}.}

\item{The known WDs with circumstellar dust disks form a broad sequence extending from the locus of naked WDs (with blue colors) to the upper right (red) corner of the color-color diagram.  The new candidate WDs with dust disks reported here (see below) follow this sequence at the red end, and broaden it at the blue end.}
\end{itemize}

Our criteria initially resulted in the selection of 
%REF-NUMBER
28 
WDs from the MS99 sample.  
%REF-NUMBER 
Seven 
of these were excluded because they are known WDs with circumstellar dust disks: 
WD~0408-041 \citep{kvl06}, 
WD~0843+516 \citep{boris12},
WD~1015+161 \citep{jfz07a}, 
WD~1116+026 \citep{jfz07a}, 
WD~1150-153 \citep{kr07},
WD~1541+650 \citep{kilic12},
and
WD~1729+371 \citep{bfj05,kvl05}.  
%REF-NUMBER 
Five 
WDs were excluded for being known central stars of planetary nebulae (CSPNs): 
WD~0558-756 \citep{hf81,rw93},
WD~0950+139 \citep{lgb89}, 
WD~1821+643 \citep{sck85}, 
WD~1958+015 \citep{ns95},
and
WD~2333+301 \citep{sn90}.  
Another 
%REF-NUMBER 
six 
were rejected for various reasons related to contamination of their photometry: 
\begin{itemize}
\item{WD~0457-103 is in a known spectroscopic binary with a bright G--K star (63~Eri; \citealt{euveI-97,bhf94}); it remains unresolved despite a high resolution {\it Hubble Space Telescope} imaging investigation \citep{bbb01}.}

\item{WD~0725+318 has a nearby source (RA=07:28:11.65, dec=+31:43:46.57, J2000), which is revealed in the SDSS images of the field as a red background galaxy that is very bright in the IR and likely contaminates the WD photometry in both 2MASS and {\it WISE}.}

\item{WD~1109-225 is in a known unresolved spectroscopic binary with a bright A star ($\beta$~Crt; \citealt{fsb91}).}

\item{WD~1201+437 is classified as a DC+dMe binary by \citet{fgj93}. 
\citet{xwd99} classify it as a quasar based on its X-ray properties; however, we note that their X-ray error circle has a $9\arcsec$ radius, so this could be a mis-identification.  In either case, we would remove it from our sample.}

\item{WD~1235+321 has a stellar profile in the {\it WISE} images that is faint and extended, while the SDSS images show two nearby faint point sources that are likely contaminating the {\it WISE} photometry.  These sources are: a star located $5.6\arcsec$ west with ${\rm i}=20.11$~mag and ${\rm z}=19.94$~mag, and a galaxy located $6.2\arcsec$ east with ${\rm i}=19.87$~mag and ${\rm z}=19.64$~mag.  There are three additional faint sources within $13\arcsec$ of the WD (two to the south, one to the southwest in the direction of the bright star located $25\arcsec$ southwest of the WD), all with ${\rm i}>21$~mag and ${\rm z}>22$~mag.}

\item{WD~1859+429 is in a wide binary with a common proper motion companion located $\approx15\arcsec$ to the northeast.  The common proper motion companion is not problematic; however, due to its proper motion, the WD is superposed on a field star in the 2MASS images.  It is likely that this star contaminates the {\it WISE} photometry.}
\end{itemize}

This leaves 
%REF-NUMBER 
10 
WDs that we classified as targets-of-interest and for which we constructed UV--IR spectral energy distributions (SEDs).
In addition to the 2MASS \citep{2mass06} All Sky Data Release Point Source Catalog and {\it WISE} All Sky Release\footnote{See the Explanatory Supplement at \url{http://wise2.ipac.caltech.edu/docs/release/allsky/expsup/}.} photometry, we utilized photometric data from the {\it Galaxy Evolution Explorer} ({\it GALEX\/}; \citealt{galex05}), the Sloan Digital Sky Survey (SDSS Data Release 7; \citealt{sdss-dr7-09}), the AAVSO Photometric All Sky Survey (APASS Data Release 6)\footnote{See \url{http://www.aavso.org/apass}.}, and the {\it Spitzer Space Telescope} Enhanced Imaging Products Source List (Cryogenic Release v2.0, January 2013)\footnote{See the Explanatory Supplement at \\ \url{http://irsa.ipac.caltech.edu/data/SPITZER/Enhanced/Imaging/docs/Spitzer\_EIP\_expsup.pdf}.}.  
For the purpose of modeling the SEDs, we converted the various photometric measurements from magnitudes into flux densities using published zero points for each photometric band 
\citep{bessell79,2mass06,sdss-dr7-09,wise10}.
All of the photometric data for the 
%REF-NUMBER
%22 
targets-of-interest plus the CSPNs and known WDs with dust disks, along with published spectral types, effective temperatures, and surface gravities of the WDs, are listed in Tables~\ref{t:phot1}--\ref{t:phot3}.

We then fit either a DA or DB model, as appropriate for the published WD type, to the UV--optical--near-IR (JHK) portion of the target SEDs, using 
grids of H- and He-rich WD cooling models (kindly provided by P.\ Bergeron) that include the {\it GALEX}, SDSS, Johnson UBVRI, 2MASS, and {\it WISE} bands \citep{bwb95,holberg06}.
The published WD temperature and $\log$~g values from Table~\ref{t:phot1} were used as initial values (a ``typical'' value of $\log$~g = 8.0 was assumed in cases for which there is no published value).
We searched for best-fitting models within the 1$\sigma$ uncertainties of the literature values of $T_{\rm eff}$ and $\log$~g.  
For WDs with no published uncertainty for $T_{\rm eff}$, we assumed $\pm2000$~K.  

Seven of the targets have an obvious excess over the WD model in the IR.
For these objects, we re-fit the entire SED using an additional circumstellar dust disk component.  
The model dust disk SED was calculated as originally devised by \citet{jura03}, following the procedure described in \citet{wired2}, with free parameters of inner radius, width, and inclination.
The minimum allowed inner disk radius was given by a conservative sublimation temperature of $T_{\rm subl}=2100$~K; in the case of our hottest WD (WD~0420+520, $T_{\rm eff}=24300$~K), we relaxed this criterion to $T_{\rm subl}=2500$~K in order to obtain the best dust disk model fit\footnote{Silicate dust, for example, is generally assumed to sublimate at $T\gtrsim1500$--$2000$~K (e.g., \citealt{pollack94,kobayashi11}); however, \citet{rafikov12} recently calculated that due to the high metal vapor pressure at the inner edge of WD dust disks, the dust grain sublimation temperatures can be 300--400~K higher than is generally assumed, and the inner edge of the disk can be superheated to as high as 2500--3500~K.}.

\begin{figure*}[tb]
\epsscale{0.90}
\plotone{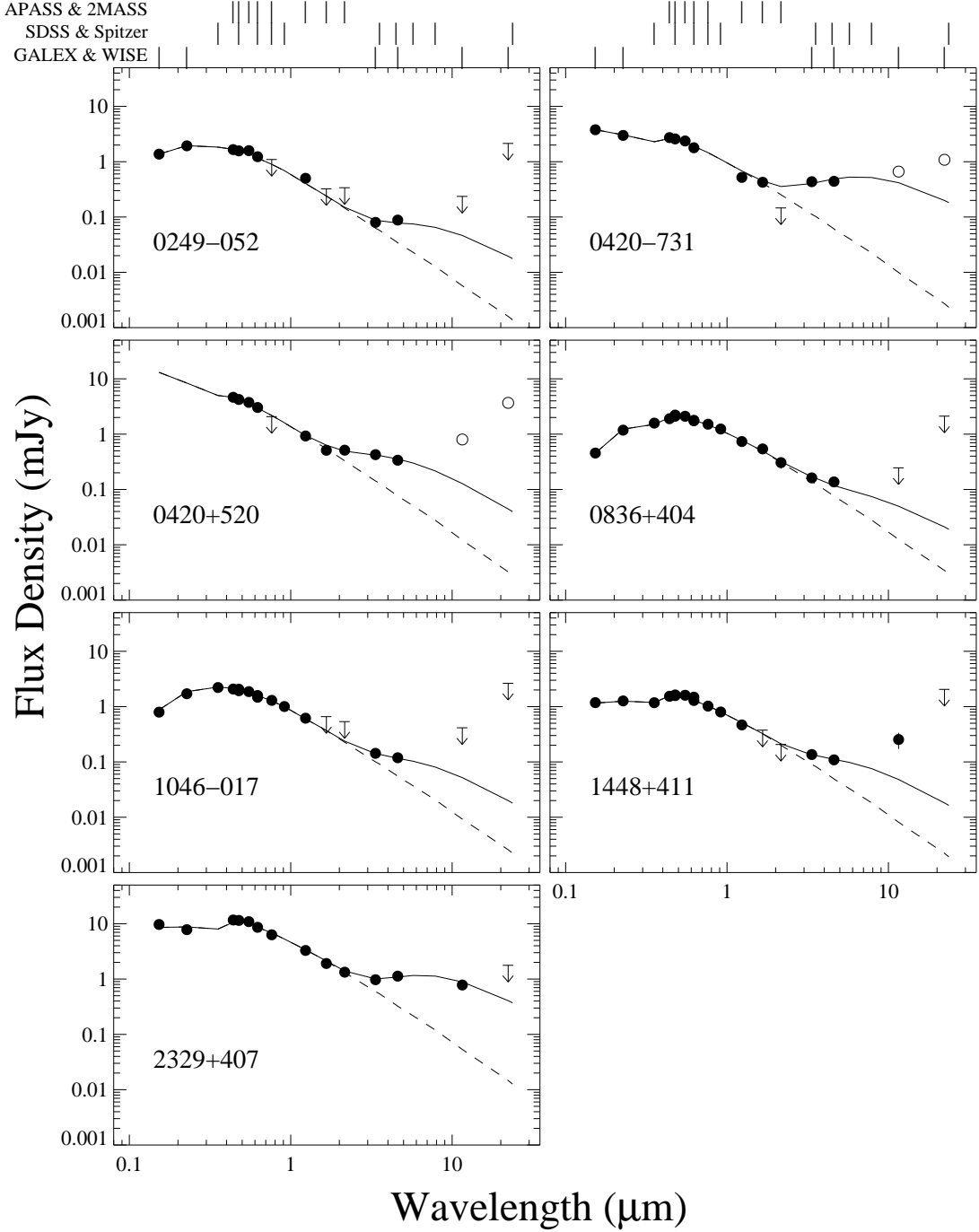}
\epsscale{1.0}
\figurenum{2}
\caption{Spectral energy distributions of the seven new WD with dust disk candidates.  Photometric values are shown as filled circles; cataloged $2\sigma$ (95\% confidence) upper limits for non-detections are shown as downward arrows.  The unfilled circles are W3 and W4 photometry that is of questionable quality and should be treated as upper limits (see Sections~\ref{s:0420-731} and \ref{s:0420+520}).  The wavelengths of the various photometric bands are indicated at the top of the figure (see Tables~\ref{t:phot1}--\ref{t:phot3} for the wavelength of each band).  The dashed line is a WD model fit, while the solid line is a combined WD + dust disk model (see Table~\ref{t:disks} for disk parameters).
The two models are generally indistinguishable shortward of $\approx2$~$\mu$m.
\label{f:sedplot1}}
\end{figure*}

As noted above, the broad 10~$\mu$m silicate emission feature that is commonly seen in the IR spectra of WDs with circumstellar dust \citep{jfz09} could contribute in the W3 band; consequently, we did not use the W3 photometry to constrain the models.
Any model that is too faint at W3 indicates the presence of a silicate emission feature or a significant amount of cool dust at large distances from the WD.  
If the model is too bright at W3, then this likely indicates that the assumed outer radius of the disk is too large.
The silicate emission feature at 18--20~$\mu$m (e.g., seen in the IR spectrum of the archetype dusty WD G29-38 = WD~2326+049; \citealt{rlv09}) is generally weaker in amplitude but broader than the 10~$\mu$m feature \citep{pp83,tvf03,jfz07b}\footnote{Laboratory experiments suggest that ``fresh'' silicates have a very large 10/20~$\mu$m flux ratio, but factors such as metal content or oxidation -- that is, age -- of the silicate compounds can increase the strength of the 20~$\mu$m feature relative to the 10~$\mu$m feature \citep{nh90}.  Dust temperature can also affect this ratio \citep{suh99}.}, and could contribute in the W4 band.
The disk inclinations are generally poorly constrained by the models.
The SEDs and model fits are shown in Figure~\ref{f:sedplot1}; the WD and dust disk model parameters are listed in Table~\ref{t:disks}.  These seven objects are our new candidate WDs with dust disks.

There are three reduced chi-squared ($\tilde{\chi}^{2}$) values listed in Table~\ref{t:disks} for each model.  These provide different measurements of the goodness of the model fit:

\begin{itemize}
\item{The parameter $\tilde{\chi}^{2}_{\rm wd}$ refers to the goodness of just the WD model component compared to only the UV--optical--near-IR (JHK) data.  In this wavelength region, we expect generally good agreement with a ``naked'' WD model even in the presence of circumstellar dust (which contributes most strongly at mid-IR and longer wavelengths). Large values of this statistic likely indicate deviations from the WD model in the UV; the {\it GALEX} data are typically among the brightest points in the SED and have very small relative errors. 
If $\tilde{\chi}^{2}_{\rm wd}$ is large because of deviations in the UV, then the values of the other two $\tilde{\chi}^{2}$ statistics will also be large, as the errors are dominated by the poor fit in the UV.
For example, removing the two UV points from the WD~2329+407 SED reduces $\tilde{\chi}^{2}_{\rm wd}$ from 107 to 2.1 and $\tilde{\chi}^{2}_{\rm disk}$ (see below) from 94 to 2.6.  
In the case of WD~1046-017, removing the UV points causes all of its $\tilde{\chi}^{2}$ values to drop to $\approx1.4$.}

\item{The parameter $\tilde{\chi}^{2}_{\rm all}$ refers to the goodness of just the WD model component compared to all of the available photometric data.  Large values (e.g., $\tilde{\chi}^{2}_{\rm all} > \tilde{\chi}^{2}_{\rm wd} > 1$) indicate the need for an additional model component in the IR.}

\item{The parameter $\tilde{\chi}^{2}_{\rm disk}$ refers to the goodness of the WD + dust disk model compared to all of the available photometric data.  If $\tilde{\chi}^{2}_{\rm disk} < \tilde{\chi}^{2}_{\rm all}$, then the model was improved by the addition of a dust disk component.
As noted above, this statistic can be large because it is dominated by a poor fit in the UV.  It can also be large because the W3 (and/or W4) data points are bright compared to the model (see above, as well as individual target notes below); for example, removing the bright W3 and W4 points from the SED of WD~0420+520 reduces $\tilde{\chi}^{2}_{\rm disk}$ from 7.8 to 0.8.}
\end{itemize}

\begin{figure*}[tb]
\epsscale{0.90}
\plotone{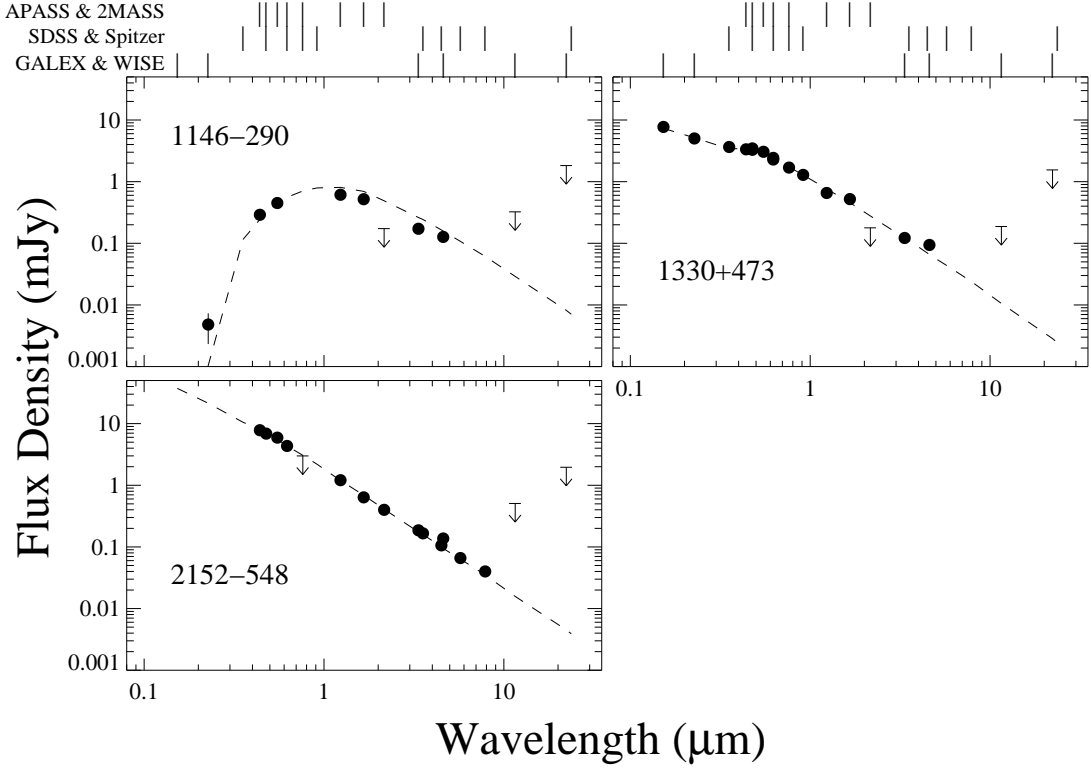}
\epsscale{1.0}
\figurenum{3}
\caption{As in Figure~\ref{f:sedplot1}, but for the three WDs selected by our target criteria that are not dust disk candidates.
\label{f:sedplot2}}
\end{figure*}

Three targets-of-interest selected by our criteria (WD~1146-290, WD~1330+473, and WD~2152-548) show no strong IR excess (see Figure~\ref{f:sedplot2}), and they are among the selected WDs with (W1$-$W2) color closest to the $+0.3$~mag criterion (see Figure~\ref{f:ccdcolor}).  The first of these is a very cool WD (see Table~\ref{t:phot1} and notes below).  It has a distinctive SED shape compared to the others in our selected sample, and its selection by our criteria was likely a by-product of its very low temperature. The latter two WDs have only very slightly elevated W2 flux densities compared to the WD model.
In particular, {\em only} the W2 point in WD~2152-548 is elevated compared to the model; the adjacent IRAC-2 point (as well as the other IRAC data) agree with the model.  Thus, the W2 value for WD~2152-548 should be treated with caution.
If real, then the IR excesses in WD~1330+473 and WD~2152-548 are very weak and could indicate that only a small amount of cool dust is present.  This situation presents similarities to the weak IR excesses found in PG~1457-086 \citep{fjz09} and HE~0106-3253 \citep{fjl10}, which are inferred to be due to narrow circumstellar rings of dust instead of full disks, as well as to the several known WDs with combined gas and dust disks (e.g., \citealt{gms06,bgm09,bgg12}).  The presence of co-mingled gas and dust disks could indicate that a significant amount of dust has been converted to gas through either sublimation \citep{mja10} or collisions (with itself or possibly with pre-existing circumstellar material; \citealt{jura08,gbf12}).

\subsection{Notes on Individual Targets}
\label{s:notes}

We have examined the publication record for each of the 
%REF-NUMBER
ten 
targets-of-interest, and briefly describe any relevant features below.
We also vetted the {\it WISE} data for each of these sources for evidence of red contaminating sources in the photometry aperture (as described in \citealt{wired2}); relevant notes are included below.

\subsubsection{WD~0249-052}
\label{s:0249}

\citet{vkn07} and \citet{lb10} do not note any atmospheric contamination (including hydrogen) in this DB WD.
In the former study, hydrogen was assumed to be absent unless a visual inspection of the optical spectrum revealed H lines (corresponding to a detection limit of H$\alpha$ equivalent width $\gtrsim300$~m\AA).  
In the latter study, it appears that hydrogen was only utilized in the model spectrum analysis if the WD had been previously identified as a hydrogen-rich helium (DBA) WD.
Additional examination of the two (somewhat noisy) spectra of this WD from the \citet{vkn07} study yields no obvious metal lines, and limits of [Ca/He] $< -8.0$ and [Mg/He] $< -6.7$ (D.\ Koester, private communication).
On the other hand, \citet{bwd11} find [H/He]$= -5.47(59)$ (but no metals) by utilizing high S/N spectra of the H$\alpha$ region.
None of these three analyses, however, conclude that metal contamination (signified by the presence of Ca absorption in the optical spectrum) is present.
We have no concerns about the quality of the {\it WISE} photometry.

\subsubsection{WD~0420-731}
\label{s:0420-731}

There is no detailed information on this WD in the literature. 
However, there is a source (WISE-J041933.70-730333.9) located $\approx22\arcsec$ northwest of the WD, which is faint in W1 and W2, but becomes much brighter in the W3 and W4 bands (slightly brighter than the WD).  This source separation is well beyond the $1.3\times({\rm FWHM}_{\rm W1})\approx7.8\arcsec$ radius, interior to which contamination of the {\em WISE} All Sky Catalog photometry can occur (see discussion in \citealt{wired2}); nonetheless, we tested for contamination from the neighboring source.  To do so, we used the IRAF\footnote{IRAF is distributed by the National Optical Astronomy Observatories, which are operated by the Association of Universities for Research in Astronomy, Inc., under cooperative agreement with the National Science Foundation.} implementation of DAOPHOT \citep{stetson87} to obtain PSF-fit photometry for the WD and the nearby source in the W3 Atlas images, using the nearby bright star WISE-J041948.50-730317.2 as a PSF template and magnitude calibrator.  We obtain 
W3$_{\rm psf}=11.69(40)$~mag for the WD and W3$_{\rm psf}=11.60(33)$~mag for the nearby source, 
in agreement with the {\it WISE} catalog values of W3~$=11.700(112)$~mag and W3~$=11.642(106)$~mag, respectively.  Nonetheless, the local background is patchy and bright in W3 and W4, so it is prudent to treat the W3 and W4 photometry as upper limits until higher resolution imaging data are available.

\subsubsection{WD~0420+520}
\label{s:0420+520}

There is no detailed information on this WD in the literature. 
The cataloged W3 and W4 flux densities for this target are quite bright, and there is no obvious point source at the position of the WD in the {\it WISE} W3 and W4 Atlas images.  
Thus, these values should be treated as upper limits.

\subsubsection{WD~0836+404}

This is a ZZ Ceti type pulsating WD \citep{vdf97}. 
\citet{farihi05} found no evidence for a low luminosity companion from a survey utilizing proper motion measurements, deep imaging, and near-IR photometry.
A limit on atmospheric metal contamination was set by \citet{zuckerman03}, at [Ca/H]$<-7.72$.

There is a bright (V=10.9~mag) star (2MASS J08401164+4015211) located $\approx43\arcsec$ east of the WD.  While this star is far enough from the WD to not pose a contamination risk for the {\it WISE} photometry, we note that a diffraction spike from the star passes near the WD in the {\it WISE} Atlas images.  Contamination warnings due to diffraction spikes are included in the {\it WISE} All Sky Catalog; such a warning was {\em not} flagged for WD~0836+404.  Nonetheless, to confirm this we performed DAOPHOT PSF-subtraction photometry on the W2 Atlas image, as described above.  
We used the nearby stars 
WISE-J084000.92+401704.4
and
WISE-J084022.86+401250.6
to construct a PSF template, and the mean photometry of the PSF stars plus several nearby stars 
(WISE-J083954.52+401509.5,
WISE-J084001.71+401415.7, 
and 
WISE-J084022.35+401424.8)
that are comparable in brightness to the WD as a magnitude calibrator.
We obtain W2$_{\rm psf}=15.29(51)$~mag for the WD, 
in agreement with its {\it WISE} catalog value of W2$=15.245(115)$~mag.  The nominal W2 PSF photometry for all 6 of the tested stars (including the WD) agrees to better than 1\% with the {\it WISE} catalog values.  So, we consider it unlikely that the nearby diffraction spike has contaminated the {\it WISE} photometry of the WD.

\subsubsection{WD~1046-017}
\label{s:1046}

This is a known DB WD and there is tentative evidence that it might be metal-rich:\ 
\citet{sak88} noted a possible weak \ion{Ca}{2}~K feature in its optical spectrum (equivalent width $<10$~m\AA), while
\citet{zuckerman10}, \citet{bwd11}, and \citet{jura12}
set limits of 
[Ca/He] $< -10.9$ (\ion{Ca}{2} equivalent width $\lesssim9$~m\AA),
[H/He] $\lesssim -6.5$,  
and
$\log(dM_{\rm metals}/dt \,\, [{\rm g}~{\rm s}^{-1}])<6.20$.
Thus, the metal-rich status of this WD remains uncertain (but unlikely).
\citet{farihi05} found no evidence for a low luminosity companion from a survey utilizing proper motion measurements, deep imaging, and near-IR photometry.
We have no concerns about the quality of the {\it WISE} photometry.

\subsubsection{WD~1146-290}

The equivalent width of H$\alpha$ in this DA WD is 5.9~\AA\ \citep{bergeron97}.
There is no other detailed information on this WD in the literature. 
We have no concerns about the quality of the {\it WISE} photometry.

\subsubsection{WD~1330+473}

No IR excess or evidence for a dust disk is noted in the near- and mid-IR photometric and spectroscopic survey of \citet{barber12}.
\citet{farihi05} found no evidence for a low luminosity companion.
We have no concerns about the quality of the {\it WISE} photometry.

\subsubsection{WD~1448+411}

There is no detailed information on this WD in the literature. 
We have no concerns about the quality of the {\it WISE} photometry.

\subsubsection{WD~2152-548}

This object was first discovered as an X-ray source in the {\it Einstein} satellite slew survey \citep{eps92}, and was 
later confirmed as a hot DA WD optical counterpart to a {\it ROSAT} X-ray source \citep{mhb95,mbb97}.
\citet{bannister03} report possible weak photospheric metal contamination in this WD from high resolution UV spectroscopic observations, and note that it is ``an object deserving of further attention.''
\citet{dickinson12} re-examined the origin of previously reported circumstellar features in 23 hot DA WDs.
While unambiguous re-detections of circumstellar material were made for eight other WDs, 
they were unable to confirm the \ion{Ca}{2} contamination reported by \citet{bannister03} for WD~2152-548.
We have no concerns about the quality of the {\it WISE} photometry.

\subsubsection{WD~2329+407}

\begin{figure*}[tb]
\epsscale{0.90}
\plotone{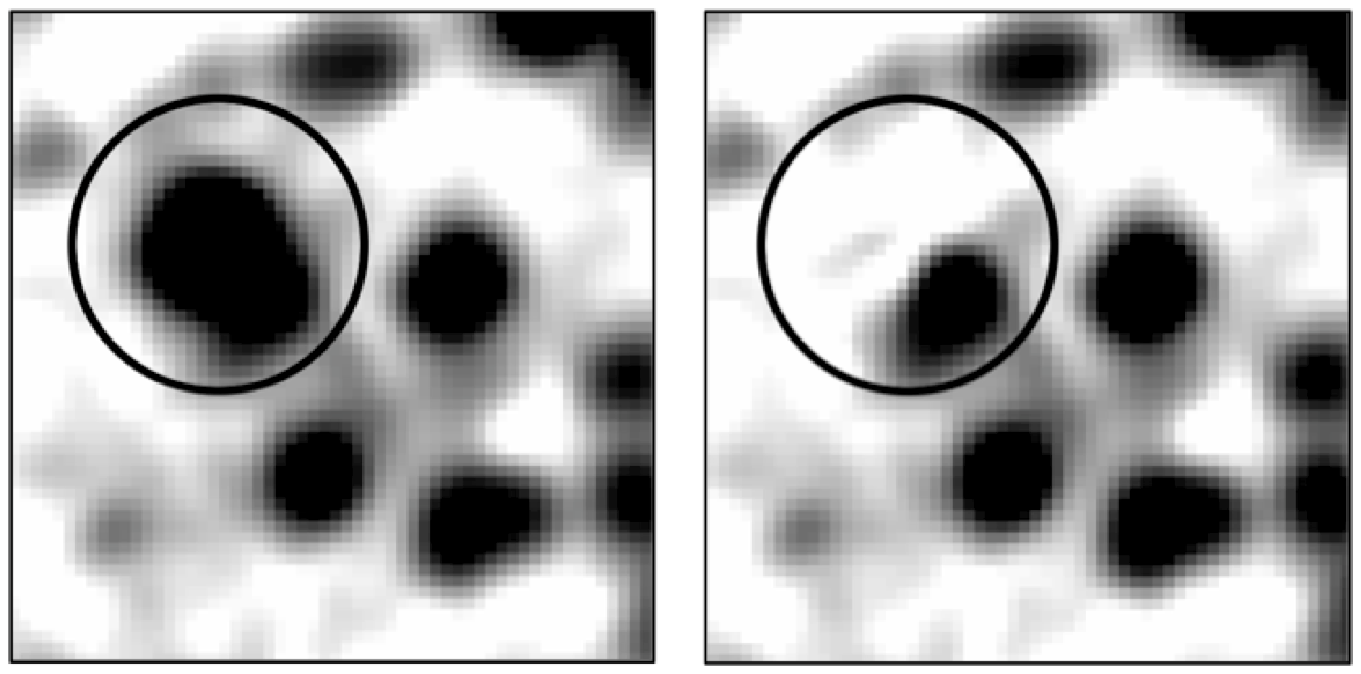}
\epsscale{1.0}
\figurenum{4}
\caption{
The {\it WISE} W1 Atlas image of WD~2329+407 before (left panel) and after (right panel) PSF-subtraction of the WD.  The circle is centered on the WD and shows the extent of the PSF radial profile fit.  The image is $\approx80\arcsec$ in both dimensions, with a plate scale of $\approx1.4\arcsec$~pixel$^{-1}$, and is oriented north up, west to the right.
This figure is representative of all of the PSF subtraction photometry performed here, as discussed in Section \ref{s:notes}.
\label{f:psfsub}}
\end{figure*}

This WD was noted as non-magnetic in the spectropolarimetric survey of \citet{ss95}.
\citet{farihi05} found no evidence for a low luminosity companion.
There is a faint field star located $\approx8\arcsec$ southwest of the WD.  Proper motion of the WD between the DSS-1 and -2 epochs shows that this object is not related to the WD.  The neighboring source is visible in the 2MASS images but is not listed in the 2MASS Point Source Catalog. The W1 Atlas image shows a slight extension to the WD PSF that is consistent with the presence of this source, but is not visible in the W2 and W3 Atlas images (the WD is not visible at all in the W4 Atlas image; the W4 photometry for the WD is an upper limit).
As with WD~0420-731, we performed DAOPHOT PSF-subtraction photometry, this time on the W1 Atlas image.  We used the nearby bright star WISE-J233143.17+410133.1 as a PSF template, and the mean photometry of several nearby stars 
(WISE-J233140.08+410100.9, WISE-J233134.97+410217.7, WISE-J233129.20+410124.4, and WISE-J233135.68+410012.5)
that are comparable in brightness to the WD as a magnitude calibrator.
The corresponding PSF-subtraction of the WD was successful (see Figure~\ref{f:psfsub}), and 
we obtain W1$_{\rm psf}=13.75(5)$~mag, 
which is in agreement with the catalog photometry of W1$=13.757(27)$~mag.  Thus, there appears to be no overt reason to be concerned about the {\it WISE} photometry of WD~2329+407.

\section{Discussion and Conclusions}

Many of the ``original'' WDs with dust disks that were discovered before 2011 have IR excesses that are detectable in the near-IR (JHK) bands.
In all but one (WD~0420+520) of the new dust disk candidates presented here, the excess emission due to dust is only apparent at wavelengths $\gtrsim3$~$\mu$m.  
It is possible that the current census of WDs with dust disks that produce an excess detectable at K-band and shorter wavelengths (e.g., using 2MASS or UKIDSS data; see \citealt{whh03,whh05,hoard07,sbd11}) is close to complete for the entire sample of known WDs (at least to the detection limits of existing near-IR all-sky surveys).
The WD dust disk candidates now being found using longer wavelength data from {\it WISE} and {\it Spitzer} are drawn from a previously underrepresented region of parameter space, in which the dust disks are overall cooler, narrower in radial extent, and/or contain fewer emitting grains.

\begin{figure*}[tb]
\epsscale{1.00}
\plotone{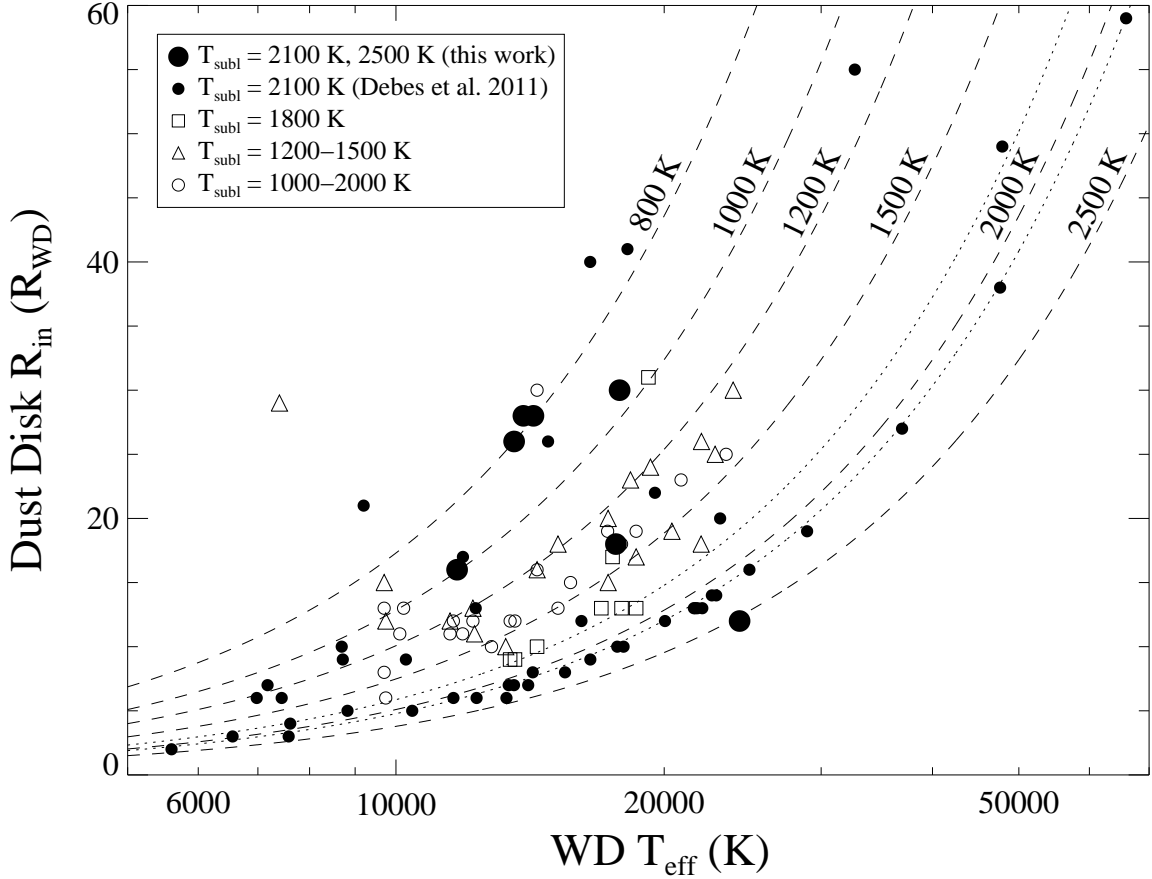}
\epsscale{1.0}
\figurenum{5}
\caption{Inner edge radius as a function of WD effective temperature for published WD dust disk models.  
The dashed lines are isotemperature contours 
that show the temperature at a given radius in the disk for a given WD temperature.
Models that are ``sublimation-limited'' (i.e., the dust located closest to the WD is constrained to be no hotter than the sublimation temperature) will lie along the isotemperature contour corresponding to the dust sublimation temperature assumed in the model.
The small filled circles are the new WD dust disk candidates from \citet{wired2} (with assumed sublimation temperature of 2100~K, shown as a dotted isotemperature contour) and the large filled circles are the new candidates from this work.  The other points are previously known WD dust disks, using published disk models and WD effective temperatures (see Table~\ref{t:pubdisks}; some WDs are plotted more than once corresponding to different published dust disk model parameters), and are symbol-coded as follows:\ 
unfilled triangles represent models with assumed sublimation temperatures of 1200--1500~K, 
unfilled squares have assumed sublimation temperature of 1800~K (shown as a dotted isotemperature contour), 
and unfilled circles have various assumed sublimation temperatures in the range 1000--2000~K.
Disk radii were rounded to integers, so some plotted points fall at radius values slightly smaller or larger than the minimum radius for the corresponding assumed sublimation temperature.  
See text for additional discussion.
\label{f:pubdisks}}
\end{figure*}

In Figure~\ref{f:pubdisks}, we have plotted the dust disk inner edge radius as a function of WD effective temperature for the seven new candidates from this paper, the candidates in \citet{wired2}, and various published dust disk models for other WDs (see Table~\ref{t:pubdisks}).
There is a direct relationship between the WD temperature and the temperature of dust at a given radius \citep{jura03},  
\begin{equation}
T_{\rm dust}(R) \propto R^{-3/4} \, T_{\rm wd},
\label{eqn:tr}
\end{equation}
which is used to plot isotemperature contours in Figure~\ref{f:pubdisks} for the model dust disks.
For an assumed dust sublimation temperature, the corresponding contour shows the minimum inner radius of the disk as a function of WD temperature; in general, the contours illustrate the temperature at a given radius in the disk for a given WD temperature.
A typical boundary condition of WD dust disk models is that no dust is allowed to be hotter than the assumed sublimation temperature; in other words, dust is not allowed at radii closer to the WD than the radius at which the sublimation temperature (according to equation~\ref{eqn:tr}) is reached.
In many cases, the hottest dust in the disk models is at the assumed sublimation temperature (disk models with inner edges at the assumed sublimation temperature are indicated in Tables~\ref{t:disks} and \ref{t:pubdisks}).  Such a disk extends inward toward the WD as far as possible -- its inner edge lies at the sublimation radius around the WD, so dust cannot survive any closer to the WD.  
On the other hand, disks that have inner edge radii corresponding to temperatures below the dust sublimation temperature could, in principle, extend inward closer to the WD but do not.
In such cases, possible reasons for the lack of dust grains close to the WD include (but are not limited to):\

\begin{itemize}
\item{Grains close to the WD might have been depleted due to a higher rate of collisions with other grains.
This might have occurred in the handful of known WDs with gas+dust disks, in which the gas and dust share spatially overlapping, but not identical, radial distributions, implying that dust sublimation alone cannot account for the presence of gas/lack of close-in dust (e.g., \citealt{bgg12}).  Additionally, objects like WD~1456+298 (G166-158) have a very weak IR excess corresponding to the presence of only cool dust (see additional discussion of this object below).} 

\item{A spinning WD with a strong magnetic field might sweep up paramagnetic or diamagnetic dust grains interior to a critical radius.  A similar process operates in the intermediate polar class of cataclysmic variable to produce a truncated gaseous accretion disk around the WD (see review of this class in \citealt{warner03}). 
Based on observations of the \ion{Ca}{2} lines during pulsations of the archetype dusty WD G29-38, \citet{tmv10} have suggested that the Ca is being accreted onto the poles of the WD, rather than equatorially, suggesting that the WD is magnetic.  
\citet{bhw07} (see the Appendix in that paper) calculated the critical surface charge on a dust grain, $Q_{\rm crit}$, such that the motion of the dust would be influenced by the WD magnetic field.  
They found that $Q_{\rm crit}$ for dust grains at the sublimation radius around a 14,200~K WD with a field strength of $B\approx25$~MG is more than an order of magnitude larger than the likely surface charge of the grains based on observations of in situ interplanetary dust grains in the Solar System.  However, they note that
the value of $Q_{\rm crit}$ is several orders of magnitude smaller than the value $Q_{\rm max}$, at which the electrostatic tensile stress in the dust grain interiors would be sufficient to fracture (i.e., destroy) the grains.
In addition, $Q_{\rm crit}\propto R^{3/2} \, B^{-1}$ (where $R$ is the distance from the center of the WD to the dust grain), leaving open the possibility that magnetic interactions could be effective around WDs with larger magnetic fields and/or cooler temperatures (allowing the dust to approach closer to the WD before sublimating).  Finally, \citet{bhw07} did not consider any intrinsic para- or diamagnetic properties of the dust itself, which could enhance interaction with the WD magnetic field.
Other than these two examples (one observational, one theoretical), there has been (to our knowledge) little exploration of this possibility.}
\end{itemize}

There is, of course, some ambiguity in the interpretation of Figure~\ref{f:pubdisks}.  
A given disk could appear to not reach the sublimation radius if the assumed sublimation temperature was higher than the true sublimation temperature for the particular species of dust in the observed disk.  
In the bulk of cases, however, the published models of dust disks extend inward to the sublimation radius even when the assumed sublimation temperature is quite high (e.g., a majority of the large sample in \citealt{wired2} with $T_{\rm subl}=2100$~K, or WD~0420+520 in this work).
Apparently, most of the currently known dusty WDs have ``hot'' disks in which the dust extends inward quite close to the WD, until it reaches the ``sublimation barrier.'' 
In contrast, five of the seven dust disk candidates presented here have maximum dust temperatures of $\lesssim1000$~K, safely below reasonable lower limits of the sublimation temperature for metallic dust.  They are, therefore, truly ``cool'' disks in which dust is depleted close to the WD and is mainly present substantially exterior to the sublimation radius.  
These disks are narrower in radial extent than an otherwise identical disk in which the inner edge extends all the way to the sublimation radius.
The apparent bias toward sublimation-limited disks does not, however, imply that ``hot'' WD dust disks are necessarily more intrinsically common than ``cool'' disks.  The ``hot'' disks are easier to find, since they produce strong IR excesses that can be detected in the near-IR.

Three of our dust disk candidates (WD~0249-052, WD~0836+404, WD~1046-017) have published optical spectroscopic studies noting the absence of atmospheric metal pollution.
One would expect to find metals in the atmosphere of a WD with an IR excess indicating a dust disk, as some of the dust will accrete onto the star.
The DB WDs (WD~0249-052, WD~1046-017) have gravitational settling times for metals in their helium-rich atmospheres that are much longer than for DA WDs; metals can persist for up to a Myr or longer after accretion ceases, making them appear as metal-rich WDs for a substantial time after the dust disk has dissipated (typical disk lifetimes are $\sim$0.03--5~Myr; \citealt{gbf12}).
While there appear to be firm constraints on the lack of accreted metals in WD~0249-052 (see Section \ref{s:0249}), the situation for WD~1046-017 is less certain (see Section \ref{s:1046}).
Possibly the most viable explanation for the presence of a dust disk around a DB WD that is not metal-polluted is that the disk is newly formed and the abundance of accreted metals on the WD is not yet in a steady state and has not reached the threshold of detection.  

On the other hand, because of the short settling times of DA WDs, essentially any process that causes the accretion of metals from the dust disk to be out of steady state could produce the phenomenon of a WD with an IR excess but no spectroscopic signature of metal pollution.
In the case of the DA WD among these three, we note that the limit on the metal contamination of WD~0836+404 is not very stringent ([Ca/H]$<-7.7$; \citealt{zuckerman03}).  This leaves some ambiguity in the conclusion that the WD has no metal pollution; 
for example, the SED of WD~0836+404 (like the other two; also see discussion above) is similar in appearance to that of G166-158.  The latter is a metal-polluted DA WD with a very weak IR excess that only becomes apparent longward of $\approx5$~$\mu$m, and likely indicates a narrow annulus of cool dust located relatively far from the WD with a large, inner, dust-free zone.
\citet{zuckerman03} measured [Ca/H]$=-9.3$ for G166-158, implying that WD~0836+404 could still be considered metal-rich with [Ca/H]$<-7.7$, even if the true Ca abundance limit is an order of magnitude lower, given that \citet{zuckerman03} assumed a somewhat higher temperature for WD~0836+404.  By the same token, WD~0836+404 ($\sim11700$~K) is hotter than G166-158 ($\sim7400$~K), so the former would require a higher Ca abundance to show lines than the latter (i.e., they could have comparable levels of Ca enrichment, but only the cooler WD might show Ca lines in its optical spectrum).
Regardless, a DA WD with a dust disk but no metal pollution might be explained by a number of scenarios, including the inner disk dust depletion and ``magnetic sweeping'' scenarios discussed above, as well as non-steady state accretion from a newly-formed dust disk (as suggested above for the case of DB WDs).  

We end with a final note of caution.
As discussed in \citet{wired2}, the large {\it WISE} PSF can defy even the most careful vetting process and allow unidentified contamination from faint, red, unresolved sources.  Thus, the seven WDs highlighted here should be considered only as dust disk {\em candidates} until independent confirmation of their IR excesses can be obtained.  This is especially true for the three WDs, discussed above, that do not appear to show the signature of metal contamination in their optical spectra.  In these cases, the existing IR data do not strongly constrain a dust disk model.  Additional longer wavelength mid-IR observations are needed to rule out the possibility that a very cool brown dwarf companion could be responsible for the observed IR excess.

\acknowledgements{
This work is based on data, data products, and other resources obtained from: 
(a) The Two Micron All Sky Survey (2MASS), a joint project of the University of Massachusetts and the Infrared Processing and Analysis Center (IPAC)/California Institute of Technology (Caltech), funded by the National Aeronautics and Space Administration (NASA) and the National Science Foundation (NSF).
(b) The American Association of Variable Star Observers (AAVSO) Photometric All-Sky Survey (APASS), funded by the Robert Martin Ayers Sciences Fund.
(c) NASA's Astrophysics Data System.  
(d) The {\it Galaxy Evolution Explorer} ({\it GALEX\/}), a NASA Small Explorer launched in April 2003 and operated for NASA by Caltech under NASA contract NAS-98034.  The {\it GALEX} data products were obtained from the Multimission Archive at the Space Telescope Science Institute (MAST). STScI is operated by the Association of Universities for Research in Astronomy, Inc., under NASA contract NAS5-26555. Support for MAST for non-HST data is provided by the NASA Office of Space Science via grant NNX09AF08G and by other grants and contracts.
(e) The NASA/IPAC Infrared Science Archive (IRSA), which is operated by the Jet Propulsion Laboratory (JPL), Caltech, under a contract with NASA, including the {\it Spitzer} Enhanced Imaging Products based on observations obtained with the {\it Spitzer Space Telescope}, which is operated by JPL, Caltech, under a contract with NASA.
(f) The SIMBAD database, operated at CDS, Strasbourg, France.
(g) The Sloan Digital Sky Survey (SDSS and SDSS-II), whose funding has been provided by the Alfred P.\ Sloan Foundation, the Participating Institutions, the NSF, the U.S.\ Department of Energy, NASA, the Japanese Monbukagakusho, the Max Planck Society, and the Higher Education Funding Council for England. The SDSS Web site is http://www.sdss.org/. The SDSS is managed by the Astrophysical Research Consortium for the Participating Institutions. The Participating Institutions are the American Museum of Natural History, Astrophysical Institute Potsdam, University of Basel, University of Cambridge, Case Western Reserve University, University of Chicago, Drexel University, Fermilab, the Institute for Advanced Study, the Japan Participation Group, Johns Hopkins University, the Joint Institute for Nuclear Astrophysics, the Kavli Institute for Particle Astrophysics and Cosmology, the Korean Scientist Group, the Chinese Academy of Sciences (LAMOST), Los Alamos National Laboratory, the Max-Planck-Institute for Astronomy (MPIA), the Max-Planck-Institute for Astrophysics (MPA), New Mexico State University, Ohio State University, University of Pittsburgh, University of Portsmouth, Princeton University, the United States Naval Observatory, and the University of Washington.
(h) The {\it Wide-field Infrared Survey Explorer} ({\it WISE\/}), which is a joint project of the University of California, Los Angeles, and JPL, Caltech, funded by NASA.

{\it Facilities:} 
\facility{2MASS}, 
\facility{AAVSO}, 
\facility{GALEX}, 
\facility{SDSS}, 
\facility{Spitzer (IRAC, MIPS)},
\facility{WISE}}

%%% BEGIN FIGURE %%%%%%%%%%%%%%%%%%%%%%%%%%%%%%%%%
%%% END FIGURE %%%%%%%%%%%%%%%%%%%%%%%%%%%%%%%%%

%%% BEGIN FIGURE %%%%%%%%%%%%%%%%%%%%%%%%%%%%%%%%%
%%% END FIGURE %%%%%%%%%%%%%%%%%%%%%%%%%%%%%%%%%

%%% BEGIN FIGURE %%%%%%%%%%%%%%%%%%%%%%%%%%%%%%%%%
%%% END FIGURE %%%%%%%%%%%%%%%%%%%%%%%%%%%%%%%%%

%%% BEGIN FIGURE %%%%%%%%%%%%%%%%%%%%%%%%%%%%%%%%%
%%% END FIGURE %%%%%%%%%%%%%%%%%%%%%%%%%%%%%%%%%

%%% BEGIN FIGURE %%%%%%%%%%%%%%%%%%%%%%%%%%%%%%%%%
%%% END FIGURE %%%%%%%%%%%%%%%%%%%%%%%%%%%%%%%%%

%%% BEGIN TABLE %%%%%%%%%%%%%%%%%%%%%%%%%%%%%%%%%
\clearpage
\begin{deluxetable}{ccccccccccc}
\tabletypesize{\scriptsize}
\setlength{\tabcolsep}{0.06in} 
\rotate
\tablecolumns{11}
\tablewidth{625pt}
\tablecaption{System Parameters and Photometry (UV, near-IR) of Selected MS99 White Dwarfs \label{t:phot1}}
\tablehead{
\colhead{ } &
\colhead{ } &
\colhead{ } &
\colhead{ } &
\colhead{ } &
\colhead{ } &
\multicolumn{2}{l}{{\it GALEX\/}:} &
\multicolumn{3}{l}{2MASS:} 
\\
\colhead{WD} &
\colhead{{\it WISE} Designation} &
\colhead{Type\tablenotemark{a}} &
\colhead{$T_{\rm eff}$} &
\colhead{$\log{g}$} &
\colhead{Refs} &
\colhead{fuv} &
\colhead{nuv} &
\colhead{J} &
\colhead{H} &
\colhead{Ks} 
\\
\colhead{ } &
\colhead{ } &
\colhead{ } &
\colhead{(K)} &
\colhead{ } &
\colhead{ } &
\colhead{(mag)} &
\colhead{(mag)} &
\colhead{(mag)} &
\colhead{(mag)} &
\colhead{(mag)} 
\\
\colhead{ } &
\colhead{ } &
\colhead{ } &
\colhead{ } &
\colhead{ } &
\colhead{ } &
\colhead{($\mu$Jy)} &
\colhead{($\mu$Jy)} &
\colhead{($\mu$Jy)} &
\colhead{($\mu$Jy)} &
\colhead{($\mu$Jy)} 
\\
\colhead{ } &
\colhead{ } &
\colhead{ } &
\colhead{ } &
\colhead{ } &
\colhead{ } &
\colhead{[0.1528~$\mu$m]} &
\colhead{[0.2271~$\mu$m]} &
\colhead{[1.235~$\mu$m]} &
\colhead{[1.662~$\mu$m]} &
\colhead{[2.159~$\mu$m]} 
}
\startdata
\multicolumn{11}{l}{New WD Dust Disk Candidates:} \\[2pt]
0249-052 & J025215.53-050231.3 &      DB &  $17700\pm548$ &   $8.16\pm0.09$ & 1     & $16.057\pm0.033$ & $15.682\pm0.016$ &     $16.256\pm0.116$ &            $>16.249$ &          $>15.735$ \\[2.0pt]   
         &                     &         &                &                 &       &      $1371\pm42$ &      $1937\pm28$ &    $501^{+52}_{-57}$ &               $<324$ &             $<339$ \\[5.5pt]
0420-731 & J041937.81-730344.3 &      DA &        \nodata &         \nodata & 2     & $14.955\pm0.019$ & $15.212\pm0.015$ &     $16.217\pm0.099$ &     $15.954\pm0.179$ &          $>16.646$ \\[2.0pt]   
         &                     &         &                &                 &       &      $3783\pm67$ &      $2987\pm40$ &    $520^{+46}_{-50}$ &    $425^{+65}_{-77}$ &             $<146$ \\[5.5pt]
0420+520 & J042415.70+521010.6 &      DA & $26142\pm3921$ &           $8.1$ & 3     &          \nodata &          \nodata &     $15.589\pm0.062$ &     $15.755\pm0.172$ &   $15.284\pm0.156$ \\[2.0pt]   
         &                     &         &                &                 &       &          \nodata &          \nodata &    $927^{+54}_{-57}$ &    $511^{+76}_{-88}$ &  $513^{+69}_{-80}$ \\[5.5pt]
0836+404 & J084007.64+401503.6 &      DA &  $11870\pm180$ &   $8.10\pm0.05$ & 4     & $17.254\pm0.048$ & $16.219\pm0.020$ &     $15.840\pm0.073$ &     $15.693\pm0.130$ &   $15.845\pm0.219$ \\[2.0pt]   
         &                     &         &                &                 &       &       $455\pm20$ &      $1181\pm22$ &    $735^{+50}_{-53}$ &    $541^{+62}_{-70}$ &  $306^{+56}_{-69}$ \\[5.5pt]
1046-017 & J104832.65-020112.3 &  DB(Z?) &  $14620\pm354$ &   $8.14\pm0.13$ & 1     & $16.651\pm0.039$ & $15.819\pm0.009$ &     $16.030\pm0.089$ &            $>15.481$ &          $>15.240$ \\[2.0pt]   
         &                     &         &                &                 &       &       $794\pm28$ &      $1708\pm13$ &    $617^{+50}_{-54}$ &               $<658$ &             $<535$ \\[5.5pt]
1448+411 & J145006.54+405533.7 &      DA &        \nodata &         \nodata & 2     & $16.221\pm0.031$ & $16.143\pm0.019$ &     $16.333\pm0.111$ &            $>16.092$ &          $>16.271$ \\[2.0pt]   
         &                     &         &                &                 &       &      $1180\pm34$ &      $1267\pm22$ &    $467^{+46}_{-51}$ &               $<375$ &             $<207$ \\[5.5pt]
2329+407 & J233135.90+410129.6 &      DA &        $15900$ &          $7.91$ & 5     & $13.933\pm0.008$ & $14.167\pm0.006$ &     $14.213\pm0.031$ &     $14.322\pm0.058$ &   $14.249\pm0.070$ \\[2.0pt]   
         &                     &         &                &                 &       &      $9698\pm71$ &      $7816\pm43$ & $3291^{+109}_{-111}$ & $1912^{+106}_{-111}$ & $1332^{+87}_{-92}$ \\[7.5pt]
\multicolumn{11}{l}{Other Selected WDs:} \\[2pt]
1146-290 & J114904.63-292150.8 &      DA &   $5770\pm140$ &          $8.00$ & 6     &          \nodata & $22.195\pm0.560$ &     $16.037\pm0.092$ &     $15.737\pm0.172$ &          $>16.462$ \\[2.0pt]   
         &                     &         &                &                 &       &          \nodata &          $5\pm3$ &    $613^{+51}_{-55}$ &    $519^{+77}_{-90}$ &             $<173$ \\[5.5pt]
1330+473 & J133236.00+470411.0 &      DA &        $22570$ &          $7.89$ & 5     & $14.181\pm0.001$ & $14.649\pm0.001$ &     $15.966\pm0.086$ &     $15.734\pm0.155$ &          $>16.434$ \\[2.0pt]   
         &                     &         &                &                 &       &       $7722\pm8$ &       $5018\pm3$ &    $655^{+51}_{-55}$ &    $521^{+70}_{-81}$ &             $<178$ \\[5.5pt]
2152-548 & J215621.35-543823.3 &      DA &  $45171\pm121$ & $7.878\pm0.009$ & 7     &         \nodata  &          \nodata &     $15.300\pm0.047$ &     $15.515\pm0.116$ &   $15.554\pm0.194$ \\[2.0pt]   
         &                     &         &                &                 &       &         \nodata  &          \nodata &   $1209^{+55}_{-57}$ &    $637^{+66}_{-73}$ &  $400^{+66}_{-79}$ \\[7.5pt]
\tablebreak
\multicolumn{11}{l}{EXCLUDED -- Known WDs with Dust Disks:} \\[2pt]
0408-041 & J041102.16-035823.7 &     DAZ &   $15414\pm45$ & $7.856\pm0.010$ & 7     & $15.580\pm0.018$ & $15.592\pm0.012$ &     $15.870\pm0.061$ &     $15.991\pm0.129$ &   $15.440\pm0.175$ \\[2.0pt]   
         &                     &         &                &                 &       &      $2128\pm35$ &      $2105\pm22$ &    $715^{+41}_{-43}$ &    $411^{+47}_{-52}$ &  $445^{+67}_{-78}$ \\[5.5pt]
0843+516 & J084702.27+512852.6 &      DA &        $23876$ &           $7.9$ & 5     & $15.132\pm0.018$ & $15.394\pm0.002$ &     $16.585\pm0.123$ &     $16.481\pm0.232$ &          $>16.543$ \\[2.0pt]   
         &                     &         &                &                 &       &      $3214\pm55$ &       $2527\pm5$ &    $370^{+40}_{-45}$ &    $262^{+51}_{-63}$ &             $<161$ \\[5.5pt]
1015+161 & J101803.75+155158.1 &     DAZ &   $19948\pm33$ & $7.925\pm0.006$ & 7     & $14.991\pm0.013$ & $15.213\pm0.006$ &     $16.131\pm0.085$ &     $16.120\pm0.222$ &   $16.003\pm0.216$ \\[2.0pt]   
         &                     &         &                &                 &       &      $3661\pm44$ &      $2983\pm17$ &    $562^{+43}_{-47}$ &    $365^{+68}_{-83}$ &  $265^{+48}_{-58}$ \\[5.5pt]
1116+026 & J111912.32+022033.9 &      DA &   $12121\pm23$ & $8.005\pm0.005$ & 7     & $15.666\pm0.007$ & $15.078\pm0.002$ &     $14.752\pm0.039$ &     $14.730\pm0.051$ &   $14.611\pm0.105$ \\[2.0pt]   
         &                     &         &                &                 &       &      $1966\pm12$ &       $3380\pm7$ &   $2003^{+79}_{-81}$ &   $1313^{+65}_{-68}$ &  $954^{+90}_{-99}$ \\[5.5pt]
1150-153 & J115315.23-153637.0 &     DAZ &    $12132(42)$ & $8.033\pm0.008$ & 7     & $17.332\pm0.056$ & $16.524\pm0.024$ &     $16.038\pm0.119$ &     $15.926\pm0.173$ &          $>16.119$ \\[2.0pt]   
         &                     &         &                &                 &       &       $424\pm22$ &       $892\pm20$ &    $613^{+64}_{-71}$ &    $436^{+65}_{-76}$ &             $<238$ \\[5.5pt]
1541+650 & J154144.90+645352.3 &      DA &  $11600\pm178$ &   $8.10\pm0.05$ & 8     & $17.002\pm0.048$ & $16.090\pm0.021$ &     $15.604\pm0.062$ &     $15.912\pm0.171$ &   $15.429\pm0.175$ \\[2.0pt]   
         &                     &         &                &                 &       &       $574\pm25$ &      $1330\pm26$ &    $914^{+53}_{-56}$ &    $442^{+65}_{-76}$ &  $449^{+67}_{-79}$ \\[5.5pt]
1729+371 & J173134.34+370518.5 &     DAZ &  $10540\pm200$ &   $8.24\pm0.04$ & 9     & $20.782\pm0.227$ & $17.257\pm0.028$ &     $16.162\pm0.093$ &            $>16.070$ &          $>15.604$ \\[2.0pt]   
         &                     &         &                &                 &       &         $18\pm4$ &       $454\pm12$ &    $547^{+46}_{-50}$ &               $<382$ &             $<382$ \\[7.5pt]
\multicolumn{11}{l}{EXCLUDED -- WD is a CSPN (central star of a planetary nebula):} \\[2pt]
0558-756 & J055702.22-754020.7 &      DO &$100000\pm15000$&     $6.5\pm0.5$ & 10    & $14.107\pm0.012$ & $14.573\pm0.009$ &     $16.403\pm0.120$ &            $>16.490$ &          $>15.818$ \\[2.0pt]  
         &                     &         &                &                 &       &      $8266\pm88$ &      $5381\pm46$ &    $438^{+46}_{-52}$ &               $<260$ &             $<314$ \\[5.5pt] 
0950+139 & J095258.96+134434.7 &      DA &        $93230$ &          $7.36$ & 11    & $13.720\pm0.001$ & $14.777\pm0.002$ &     $16.518\pm0.097$ &     $15.945\pm0.157$ &   $16.099\pm0.258$ \\[2.0pt]   
         &                     &         &                &                 &       &     $11807\pm15$ &       $4458\pm6$ &    $394^{+34}_{-37}$ &    $429^{+58}_{-67}$ &  $242^{+51}_{-65}$ \\[5.5pt]
1821+643 & J182152.09+642153.7 &     DOZ &       $140000$ &           $6.1$ & 12,13 & $12.669\pm0.001$ & $13.292\pm0.001$ &     $15.784\pm0.076$ &     $15.996\pm0.190$ &          $>15.468$ \\[2.0pt]   
         &                     &         &                &                 &       &     $31086\pm21$ &     $17502\pm11$ &    $774^{+54}_{-58}$ &    $409^{+66}_{-79}$ &             $<433$ \\[5.5pt]
\tablebreak
1958+015 & J200039.34+014341.6 &  PG1159 &        $58900$ &         \nodata & 14    & $14.688\pm0.008$ & $15.286\pm0.005$ &     $15.428\pm0.065$ &     $14.830\pm0.074$ &   $14.603\pm0.106$ \\[2.0pt]   
         &                     &         &                &                 &       &      $4840\pm37$ &      $2789\pm14$ &   $1075^{+65}_{-69}$ &   $1198^{+82}_{-88}$ & $961^{+91}_{-100}$ \\[5.5pt]
2333+301 & J233553.29+302805.6 &     DOZ &       $150000$ &           $7.0$ & 15,16 &          \nodata & $14.603\pm0.002$ &     $16.697\pm0.136$ &            $>16.995$ &          $>17.164$ \\[2.0pt]   
         &                     &         &                &                 &       &          \nodata &      $5235\pm10$ &    $334^{+40}_{-45}$ &               $<163$ &              $<91$ 
\enddata
\tablenotetext{a}{From MS99 and MSonline.}
\tablecomments{Upper limits are $2\sigma$ (95\% confidence) levels. 
References: 
(1) \citet{bwd11}, 
(2) no published $T_{\rm eff}$ or $\log{g}$ values, 
(3) \citet{sion88}, 
(4) \citet{lb10}, 
(5) \citet{holberg06}, 
(6) \citet{bergeron97},
(7) \citet{koester09}, 
(8) \citet{liebert05}, 
(9) \citet{zuckerman07}, 
(10) \citet{rauch94},
(11) \citet{gian10}, 
(12) \citet{bradley00}, 
(13) \citet{rauch95}, 
(14) \citet{kwitter89},
(15) \citet{chu09}, 
(16) \citet{napi93}
}
\end{deluxetable}

\clearpage
\begin{deluxetable}{ccccccccccc}
\tabletypesize{\scriptsize}
\setlength{\tabcolsep}{0.035in} 
\rotate
\tablecolumns{11}
\tablewidth{630pt}
\tablecaption{Optical Photometry of Selected MS99 White Dwarfs \label{t:phot2}}
\tablehead{
\colhead{ } &
\multicolumn{5}{l}{SDSS:} &
\multicolumn{5}{l}{APASS:} 
\\
\colhead{WD} &
\colhead{u} &
\colhead{g} &
\colhead{r} &
\colhead{i} &
\colhead{z} &
\colhead{B} &
\colhead{V} &
\colhead{g} &
\colhead{r} &
\colhead{i} 
\\
\colhead{ } &
\colhead{(mag)} &
\colhead{(mag)} &
\colhead{(mag)} &
\colhead{(mag)} &
\colhead{(mag)} &
\colhead{(mag)} &
\colhead{(mag)} &
\colhead{(mag)} &
\colhead{(mag)} &
\colhead{(mag)} 
\\
\colhead{ } &
\colhead{($\mu$Jy)} &
\colhead{($\mu$Jy)} &
\colhead{($\mu$Jy)} &
\colhead{($\mu$Jy)} &
\colhead{($\mu$Jy)} &
\colhead{($\mu$Jy)} &
\colhead{($\mu$Jy)} &
\colhead{($\mu$Jy)} &
\colhead{($\mu$Jy)} &
\colhead{($\mu$Jy)} 
\\
\colhead{ } &
\colhead{[0.3543~$\mu$m]} &
\colhead{[0.4770~$\mu$m]} &
\colhead{[0.6231~$\mu$m]} &
\colhead{[0.7625~$\mu$m]} &
\colhead{[0.9134~$\mu$m]} &
\colhead{[0.44~$\mu$m]} &
\colhead{[0.55~$\mu$m]} &
\colhead{[0.4770~$\mu$m]} &
\colhead{[0.6231~$\mu$m]} &
\colhead{[0.7625~$\mu$m]} 
}
\startdata
\multicolumn{10}{l}{New WD Dust Disk Candidates:} \\[2pt]
%                      {sdss-u} &             {sdss-g} &                   {sdss-r} &             {sdss-i} &          {sdss-z} &             {apass-B} &             {apass-V} &              {apass-g} &               {apass-r} & {apass-i} 
0249-052 &              \nodata &              \nodata &                    \nodata &              \nodata &           \nodata &      $15.993\pm0.006$ &      $15.943\pm0.089$ &       $15.911\pm0.006$ &        $16.174\pm0.068$ &            $>16.300$ \\[2.0pt]
         &              \nodata &              \nodata &                    \nodata &              \nodata &           \nodata &           $1655\pm51$ &  $1586^{+134}_{-144}$ &            $1569\pm79$ &     $1231^{+97}_{-101}$ &              $<1097$ \\[5.5pt]
0420-731 &              \nodata &              \nodata &                    \nodata &              \nodata &           \nodata &      $15.450\pm0.077$ &      $15.507\pm0.023$ &       $15.371\pm0.039$ &        $15.770\pm0.017$ &              \nodata \\[2.0pt]
         &              \nodata &              \nodata &                    \nodata &              \nodata &           \nodata &  $2729^{+204}_{-217}$ &           $2370\pm87$ &   $2580^{+158}_{-160}$ &             $1787\pm94$ &              \nodata \\[5.5pt]
0420+520 &              \nodata &              \nodata &                    \nodata &              \nodata &           \nodata &      $14.873\pm0.062$ &      $15.009\pm0.094$ &       $14.833\pm0.089$ &        $15.195\pm0.043$ &            $>15.593$ \\[2.0pt]
         &              \nodata &              \nodata &                    \nodata &              \nodata &           \nodata &  $4642^{+293}_{-306}$ &  $3750^{+331}_{-357}$ &   $4235^{+395}_{-419}$ &    $3034^{+192}_{-195}$ &              $<2074$ \\[5.5pt]
0836+404 &     $15.940\pm0.013$ &     $15.537\pm0.015$ &           $15.775\pm0.019$ &     $15.935\pm0.016$ &  $16.138\pm0.016$ &      $15.842\pm0.072$ &      $15.636\pm0.042$ &       $15.591\pm0.096$ &        $15.795\pm0.091$ &            $>16.030$ \\[2.0pt]
         &          $1585\pm82$ &         $2214\pm115$ &            $1.778\pm0.094$ &          $1513\pm79$ &       $1239\pm65$ &  $1902^{+135}_{-142}$ &  $2105^{+102}_{-104}$ &   $2107^{+207}_{-222}$ &    $1746^{+165}_{-176}$ &              $<1387$ \\[5.5pt]
1046-017 &     $15.578\pm0.020$ &     $15.614\pm0.020$ &           $15.898\pm0.015$ &     $16.104\pm0.014$ &  $16.365\pm0.025$ &      $15.745\pm0.023$ &      $15.766\pm0.036$ &       $15.683\pm0.043$ &        $15.977\pm0.057$ &              \nodata \\[2.0pt]
         & $2211^{+118}_{-119}$ &         $2062\pm110$ &  $1.588^{+0.082}_{-0.083}$ &          $1296\pm67$ &       $1004\pm55$ &   $2079^{+76}_{-77}$  &    $1867^{+83}_{-84}$ &   $1936^{+123}_{-125}$ &    $1476^{+106}_{-109}$ &              \nodata \\[5.5pt]
1448+411 &     $16.265\pm0.018$ &     $15.873\pm0.018$ &           $16.124\pm0.011$ &     $16.361\pm0.015$ &  $16.613\pm0.018$ &      $16.073\pm0.033$ &      $15.931\pm0.024$ &       $15.903\pm0.021$ &        $15.974\pm0.147$ &            $>16.486$ \\[2.0pt]
         &          $1175\pm62$ &          $1625\pm86$ &            $1.290\pm0.066$ &          $1022\pm53$ &        $800\pm42$ &   $1537^{+65}_{-66}$  &    $1604^{+59}_{-60}$ &            $1581\pm85$ &    $1481^{+202}_{-227}$ &               $<924$ \\[5.5pt]
2329+407 &              \nodata &              \nodata &                    \nodata &              \nodata &           \nodata &      $13.870\pm0.032$ &      $13.849\pm0.027$ &       $13.750\pm0.034$ &        $14.061\pm0.042$ &     $14.384\pm0.078$ \\[2.0pt]
         &              \nodata &              \nodata &                    \nodata &              \nodata &           \nodata & $11694^{+488}_{-496}$ & $10915^{+422}_{-427}$ &  $11482^{+676}_{-682}$ &    $8622^{+542}_{-550}$ & $6316^{+540}_{-567}$ \\[8.5pt]
\multicolumn{10}{l}{Other Selected WDs:} \\[2pt]
1146-290 &              \nodata &              \nodata &                    \nodata &              \nodata &           \nodata &        $17.90\pm0.05$ &        $17.30\pm0.03$ &                \nodata &                 \nodata &              \nodata \\[2.0pt]
         &              \nodata &              \nodata &                    \nodata &              \nodata &           \nodata &            $290\pm20$ &            $450\pm20$ &                \nodata &                 \nodata &              \nodata \\[5.5pt]
1330+473 &     $15.035\pm0.023$ &     $15.052\pm0.019$ &           $15.437\pm0.020$ &     $15.815\pm0.024$ &  $16.096\pm0.017$ &      $15.229\pm0.024$ &      $15.233\pm0.025$ &       $15.108\pm0.040$ &        $15.503\pm0.108$ &            $>15.776$ \\[2.0pt]
         &         $3648\pm199$ & $3460^{+183}_{-184}$ &            $2.428\pm0.130$ &   $1691^{+92}_{-93}$ &       $1287\pm68$ &  $3345^{+124}_{-125}$ &  $3051^{+115}_{-116}$ &   $3287^{+203}_{-206}$ &    $2285^{+245}_{-265}$ &              $<1777$ \\[5.5pt]
2152-548 &              \nodata &              \nodata &                    \nodata &              \nodata &           \nodata &      $14.306\pm0.016$ &      $14.512\pm0.037$ &       $14.306\pm0.016$ &        $14.805\pm0.021$ &            $>15.210$ \\[2.0pt]
         &              \nodata &              \nodata &                    \nodata &              \nodata &           \nodata &  $7826^{+261}_{-262}$ &  $5927^{+266}_{-271}$ &   $6881^{+359}_{-360}$ &    $4345^{+233}_{-234}$ &              $<2992$ \\[8.5pt]
\tablebreak
\multicolumn{11}{l}{EXCLUDED -- Known WDs with Dust Disks:} \\[2pt]
0408-041 &              \nodata &              \nodata &                    \nodata &              \nodata &           \nodata &      $15.640\pm0.054$ &      $15.513\pm0.089$ &       $15.512\pm0.058$ &        $15.794\pm0.066$ &     $15.944\pm0.056$ \\[2.0pt]
         &              \nodata &              \nodata &                    \nodata &              \nodata &           \nodata &  $2291^{+131}_{-136}$ &  $2357^{+198}_{-213}$ &   $2266^{+164}_{-168}$ &    $1748^{+135}_{-140}$ & $1501^{+107}_{-109}$ \\[5.5pt]
0843+516 &     $15.824\pm0.032$ &     $15.876\pm0.022$ &           $16.234\pm0.014$ &     $16.563\pm0.016$ &  $16.862\pm0.018$ &      $15.989\pm0.079$ &      $16.044\pm0.046$ &       $15.809\pm0.079$ &        $16.367\pm0.201$ &              \nodata \\[2.0pt]
         & $1764^{+102}_{-103}$ &          $1621\pm88$ &            $1.165\pm0.060$ &           $849\pm44$ &        $636\pm34$ &  $1661^{+127}_{-135}$ &    $1445^{+74}_{-76}$ &   $1724^{+149}_{-156}$ &    $1031^{+182}_{-216}$ &              \nodata \\[5.5pt]
1015+161 &     $15.546\pm0.021$ &     $15.465\pm0.025$ &           $15.800\pm0.027$ &     $16.083\pm0.016$ &  $16.382\pm0.026$ &      $15.596\pm0.043$ &      $15.607\pm0.061$ &       $15.499\pm0.040$ &        $15.834\pm0.077$ &            $>16.195$ \\[2.0pt]
         &         $2278\pm122$ & $2365^{+130}_{-131}$ &  $1.738^{+0.097}_{-0.098}$ &          $1320\pm69$ &        $989\pm55$ &  $2385^{+117}_{-120}$ &  $2162^{+135}_{-141}$ &   $2293^{+142}_{-144}$ &    $1684^{+143}_{-150}$ &             $<1.191$ \\[5.5pt]
1116+026 &     $14.942\pm0.013$ &     $14.536\pm0.024$ &           $14.702\pm0.019$ &     $14.956\pm0.020$ &  $15.194\pm0.013$ &      $14.734\pm0.032$ &      $14.617\pm0.021$ &       $14.558\pm0.006$ &        $14.744\pm0.014$ &     $14.957\pm0.052$ \\[2.0pt]
         &         $3973\pm205$ & $5568^{+304}_{-305}$ &  $4.777^{+0.254}_{-0.255}$ & $3730^{+199}_{-200}$ &      $2954\pm152$ &  $5277^{+220}_{-224}$ &  $5380^{+191}_{-192}$ &           $5455\pm275$ &            $4596\pm238$ & $3726^{+255}_{-261}$ \\[5.5pt]
1150-153 &              \nodata &              \nodata &                    \nodata &              \nodata &           \nodata &      $16.110\pm0.059$ &      $15.963\pm0.066$ &              $>15.830$ &               $>16.089$ &              \nodata \\[2.0pt]
         &              \nodata &              \nodata &                    \nodata &              \nodata &           \nodata &    $1486^{+90}_{-94}$ &  $1557^{+103}_{-108}$ &                $<1691$ &                 $<1332$ &              \nodata \\[5.5pt]
1541+650 &     $15.952\pm0.020$ &     $15.550\pm0.021$ &           $15.663\pm0.018$ &     $15.823\pm0.014$ &  $16.108\pm0.014$ &      $15.562\pm0.016$ &      $15.523\pm0.006$ &       $15.412\pm0.090$ &        $15.621\pm0.035$ &              \nodata \\[2.0pt]
         &          $1567\pm84$ &         $2187\pm117$ &            $1.972\pm0.104$ &          $1678\pm87$ &       $1273\pm66$ &           $2461\pm82$ &           $2336\pm71$ &   $2484^{+234}_{-248}$ &    $2049^{+122}_{-123}$ &              \nodata \\[5.5pt]
1729+371 &              \nodata &              \nodata &                    \nodata &              \nodata &           \nodata &               \nodata &               \nodata &                \nodata &                 \nodata &              \nodata \\[2.0pt]
         &              \nodata &              \nodata &                    \nodata &              \nodata &           \nodata &               \nodata &               \nodata &                \nodata &                 \nodata &              \nodata \\[8.5pt]
\multicolumn{11}{l}{EXCLUDED -- WD is a CSPN (central star of a planetary nebula):} \\[2pt]
0558-756 &              \nodata &              \nodata &                    \nodata &              \nodata &           \nodata &      $16.207\pm0.068$ &      $15.303\pm0.082$ &       $15.667\pm0.011$ &        $14.946\pm0.040$ &     $14.777\pm0.016$ \\[2.0pt]
         &              \nodata &              \nodata &                    \nodata &              \nodata &           \nodata &    $1359^{+92}_{-97}$ &           $2336\pm71$ &           $1964\pm100$ &    $3816^{+236}_{-239}$ &         $4398\pm230$ \\[5.5pt]
0950+139 &     $15.250\pm0.068$ &     $15.640\pm0.035$ &           $16.207\pm0.018$ &     $16.515\pm0.034$ &  $16.862\pm0.028$ &      $15.793\pm0.027$ &      $15.930\pm0.078$ &       $15.778\pm0.038$ &        $16.153\pm0.080$ &              \nodata \\[2.0pt]
         & $2992^{+236}_{-245}$ & $2014^{+119}_{-121}$ &            $1.195\pm0.063$ &    $887^{+52}_{-53}$ &        $636\pm36$ &    $1990^{+77}_{-78}$ &  $1605^{+121}_{-129}$ &   $1773^{+108}_{-109}$ &    $1256^{+109}_{-115}$ &              \nodata \\[5.5pt]
1821+643 &     $14.257\pm0.027$ &     $14.732\pm0.020$ &           $15.258\pm0.022$ &     $15.702\pm0.015$ &  $16.075\pm0.023$ &               \nodata &               \nodata &                \nodata &                 \nodata &              \nodata \\[2.0pt]
         & $7468^{+416}_{-418}$ & $4648^{+248}_{-249}$ &            $2.862\pm0.155$ &          $1876\pm98$ &       $1312\pm71$ &               \nodata &               \nodata &                \nodata &                 \nodata &              \nodata \\[5.5pt]
\tablebreak
1958+015 &              \nodata &              \nodata &                    \nodata &              \nodata &           \nodata &               \nodata &               \nodata &                \nodata &                 \nodata &              \nodata \\[2.0pt]
         &              \nodata &              \nodata &                    \nodata &              \nodata &           \nodata &               \nodata &               \nodata &                \nodata &                 \nodata &              \nodata \\[5.5pt]
2333+301 &              \nodata &              \nodata &                    \nodata &              \nodata &           \nodata &               \nodata &               \nodata &                \nodata &                 \nodata &              \nodata \\[2.0pt]
         &              \nodata &              \nodata &                    \nodata &              \nodata &           \nodata &               \nodata &               \nodata &                \nodata &                 \nodata &              \nodata 
\enddata
\tablenotetext{a}{WD~1146-290 is too faint for detection by APASS; the tabulated BV photometry is from \citet{bergeron97}.}
\tablecomments{Upper limits are $2\sigma$ (95\% confidence) levels.}
\end{deluxetable}

\clearpage
\begin{deluxetable}{ccccccccccc}
\tabletypesize{\scriptsize}
\setlength{\tabcolsep}{0.10in} 
\rotate
\tablecolumns{11}
\tablewidth{625pt}
\tablecaption{Mid-infrared Photometry of Selected MS99 White Dwarfs \label{t:phot3}}
\tablehead{
\colhead{ } &
\multicolumn{4}{l}{{\it WISE\/}:} &
\multicolumn{6}{l}{{\it Spitzer\/}:} 
\\
\colhead{WD} &
\colhead{W1} &
\colhead{W2} &
\colhead{W3} &
\colhead{W4} &
\colhead{Types\tablenotemark{a}} &
\colhead{IRAC-1} &
\colhead{IRAC-2} &
\colhead{IRAC-3} &
\colhead{IRAC-4} &
\colhead{MIPS-1} 
\\
\colhead{ } &
\colhead{(mag)} &
\colhead{(mag)} &
\colhead{(mag)} &
\colhead{(mag)} &
\colhead{ } &
\colhead{(mag)} &
\colhead{(mag)} &
\colhead{(mag)} &
\colhead{(mag)} &
\colhead{(mag)} 
\\
\colhead{ } &
\colhead{($\mu$Jy)} &
\colhead{($\mu$Jy)} &
\colhead{($\mu$Jy)} &
\colhead{($\mu$Jy)} &
\colhead{ } &
\colhead{($\mu$Jy)} &
\colhead{($\mu$Jy)} &
\colhead{($\mu$Jy)} &
\colhead{($\mu$Jy)} &
\colhead{($\mu$Jy)} 
\\
\colhead{ } &
\colhead{[3.35~$\mu$m]} &
\colhead{[4.60~$\mu$m]} &
\colhead{[11.56~$\mu$m]} &
\colhead{[22.24~$\mu$m]} &
\colhead{ } &
\colhead{[3.550~$\mu$m]} &
\colhead{[4.493~$\mu$m]} &
\colhead{[5.731~$\mu$m]} &
\colhead{[7.872~$\mu$m]} &
\colhead{[23.68~$\mu$m]} 
}
\startdata
\multicolumn{11}{l}{New WD Dust Disk Candidates:} \\[2pt]
%                       {W1} &               {W2} &                  {W3} &                       {W4} & {Types} &   {IRAC-1} &   {IRAC-2} &   {IRAC-3} &   {IRAC-4} & {MIPS-1} 
0249-052 &  $16.468\pm0.078$ &   $15.732\pm0.149$ &             $>12.819$ &                   $>8.981$ &         &    \nodata &    \nodata &    \nodata &    \nodata &      \nodata \\[2.0pt]
         &          $80\pm6$ &   $88^{+11}_{-13}$ &                $<236$ &                    $<2138$ &   00000 &    \nodata &    \nodata &    \nodata &    \nodata &      \nodata \\[5.5pt]
0420-731 &  $14.631\pm0.029$ &   $13.974\pm0.033$ &      $11.700\pm0.112$\tablenotemark{b} &            $9.718\pm0.517$\tablenotemark{b} &         &    \nodata &    \nodata &    \nodata &    \nodata &      \nodata \\[2.0pt]
         &        $435\pm13$ &         $442\pm15$ &     $662^{+66}_{-73}$\tablenotemark{b} &       $1084^{+411}_{-662}$\tablenotemark{b} &   00000 &    \nodata &    \nodata &    \nodata &    \nodata &      \nodata \\[5.5pt]
0420+520 &  $14.653\pm0.039$ &   $14.265\pm0.060$ &      $11.492\pm0.163$\tablenotemark{b} &            $8.392\pm0.235$\tablenotemark{b} &         &    \nodata &    \nodata &    \nodata &    \nodata &      \nodata \\[2.0pt]
         &        $426\pm17$ &  $338^{+19}_{-20}$ &   $801^{+112}_{-130}$\tablenotemark{b} &       $3678^{+718}_{-890}$\tablenotemark{b} &   00000 &    \nodata &    \nodata &    \nodata &    \nodata &      \nodata \\[5.5pt]
0836+404 &  $15.700\pm0.057$ &   $15.245\pm0.115$ &             $>12.776$ &                   $>8.991$ &         &    \nodata &    \nodata &    \nodata &    \nodata &      \nodata \\[2.0pt]
         &         $162\pm9$ &  $137^{+14}_{-15}$ &                $<246$ &                    $<2118$ &   00000 &    \nodata &    \nodata &    \nodata &    \nodata &      \nodata \\[5.5pt]
1046-017 &  $15.836\pm0.070$ &   $15.402\pm0.123$ &             $>12.208$ &                   $>8.753$ &         &    \nodata &    \nodata &    \nodata &    \nodata &      \nodata \\[2.0pt]
         &  $143^{+9}_{-10}$ &  $119^{+13}_{-14}$ &                $<414$ &                    $<2637$ &   00000 &    \nodata &    \nodata &    \nodata &    \nodata &      \nodata \\[5.5pt]
1448+411 &  $15.890\pm0.044$ &   $15.497\pm0.082$ &      $12.740\pm0.302$ &                   $>9.027$ &         &    \nodata &    \nodata &    \nodata &    \nodata &      \nodata \\[2.0pt]
         &         $136\pm6$ &    $109^{+8}_{-9}$ &     $254^{+62}_{-82}$ &                    $<2049$ &   00000 &    \nodata &    \nodata &    \nodata &    \nodata &      \nodata \\[5.5pt]
2329+407 &  $13.757\pm0.027$ &   $12.956\pm0.027$ &      $11.520\pm0.184$ &                   $>9.184$ &         &    \nodata &    \nodata &    \nodata &    \nodata &      \nodata \\[2.0pt]
         &        $973\pm28$ & $1129^{+32}_{-33}$ &   $781^{+122}_{-145}$ &                    $<1773$ &   00000 &    \nodata &    \nodata &    \nodata &    \nodata &      \nodata \\[5.5pt]
\multicolumn{11}{l}{Other Selected WDs:} \\[2pt]
1146-290 &  $15.640\pm0.054$ &   $15.327\pm0.121$ &             $>12.474$ &                   $>9.154$ &         &    \nodata &    \nodata &    \nodata &    \nodata &      \nodata \\[2.0pt]
         &         $172\pm9$ &  $127^{+14}_{-15}$ &                $<324$ &                    $<1823$ &   00000 &    \nodata &    \nodata &    \nodata &    \nodata &      \nodata \\[5.5pt]
1330+473 &  $16.011\pm0.058$ &   $15.660\pm0.122$ &             $>13.071$ &                   $>9.330$ &         &    \nodata &    \nodata &    \nodata &    \nodata &      \nodata \\[2.0pt]
         &         $122\pm7$ &   $94^{+10}_{-11}$ &                $<187$ &                    $<1550$ &   00000 &    \nodata &    \nodata &    \nodata &    \nodata &      \nodata \\[5.5pt]
2152-548 &  $15.555\pm0.057$ &   $15.249\pm0.121$ &             $>11.990$ &                   $>9.075$ &         &    \nodata &    \nodata &    \nodata &    \nodata &      \nodata \\[2.0pt]
         &        $186\pm10$ &  $137^{+15}_{-16}$ &                $<507$ &                    $<1960$ &   11110 &  $166\pm1$ &  $106\pm1$ &   $66\pm1$ &   $40\pm2$ &      \nodata \\[5.5pt]
\tablebreak
\multicolumn{11}{l}{EXCLUDED -- Known WDs with Dust Disks:} \\[2pt]
0408-041 &  $13.889\pm0.028$ &   $13.022\pm0.030$ &      $11.565\pm0.183$ &                   $>9.207$ &         &    \nodata &    \nodata &    \nodata &    \nodata &      \nodata \\[2.0pt]
         & $861^{+25}_{-26}$ & $1062^{+33}_{-34}$ &   $749^{+117}_{-138}$ &                    $<1736$ &   11111 & $1024\pm1$ & $1194\pm2$ & $1194\pm2$ & $1118\pm3$ &   $312\pm21$ \\[5.5pt]
0843+516 &  $15.898\pm0.061$ &   $15.301\pm0.108$ &             $>12.159$ &                   $>8.815$ &         &    \nodata &    \nodata &    \nodata &    \nodata &      \nodata \\[2.0pt]
         &         $135\pm8$ &  $130^{+12}_{-14}$ &                $<434$ &                    $<2490$ &   11110 &  $137\pm1$ &  $127\pm1$ &  $103\pm2$ &  $162\pm5$ &      \nodata \\[5.5pt]
1015+161 &  $15.521\pm0.048$ &   $14.989\pm0.083$ &             $>12.252$ &            $9.046\pm0.438$ &         &    \nodata &    \nodata &    \nodata &    \nodata &      \nodata \\[2.0pt]
         &         $192\pm9$ &  $174^{+13}_{-14}$ &                $<398$ &      $2014^{+669}_{-1001}$ &   11113 &  $189\pm1$ &  $160\pm1$ &  $139\pm2$ &  $124\pm4$ &        $<70$ \\[5.5pt]
1116+026 &  $14.215\pm0.031$ &   $13.788\pm0.046$ &      $12.417\pm0.541$ &                   $>8.827$ &         &    \nodata &    \nodata &    \nodata &    \nodata &      \nodata \\[2.0pt]
         & $638^{+20}_{-21}$ &  $525^{+23}_{-24}$ &   $342^{+134}_{-221}$ &                    $<2464$ &   11112 &  $589\pm1$ &  $522\pm1$ &  $480\pm2$ &  $459\pm3$ &  $1558\pm32$ \\[5.5pt]
1150-153 &  $14.681\pm0.036$ &   $13.736\pm0.044$ &      $12.128\pm0.386$ &                   $>8.645$ &         &    \nodata &    \nodata &    \nodata &    \nodata &      \nodata \\[2.0pt]
         &        $415\pm15$ &  $550^{+23}_{-24}$ &   $446^{+134}_{-191}$ &                    $<2913$ &   11110 &  $516\pm2$ &  $586\pm2$ &  $614\pm2$ &  $601\pm4$ &      \nodata \\[5.5pt]
1541+650 &  $14.638\pm0.027$ &   $13.803\pm0.030$ &      $12.928\pm0.306$ &                   $>9.374$ &         &    \nodata &    \nodata &    \nodata &    \nodata &      \nodata \\[2.0pt]
         & $432^{+12}_{-13}$ &         $517\pm16$ &     $214^{+53}_{-70}$ &                    $<1489$ &   00000 &    \nodata &    \nodata &    \nodata &    \nodata &      \nodata \\[5.5pt]
1729+371 &  $14.965\pm0.034$ &   $14.144\pm0.042$ &      $11.703\pm0.140$ &                   $>8.955$ &         &    \nodata &    \nodata &    \nodata &    \nodata &      \nodata \\[2.0pt]
         &        $320\pm11$ &  $378^{+15}_{-16}$ &     $660^{+80}_{-91}$ &                    $<2190$ &   11111 &  $371\pm1$ &  $392\pm1$ &  $422\pm1$ &  $643\pm2$ &    $300\pm9$ \\[5.5pt]
\multicolumn{11}{l}{EXCLUDED -- WD is a CSPN (central star of a planetary nebula):} \\[2pt]
0558-756 &  $16.877\pm0.137$ &   $14.677\pm0.052$ &      $10.394\pm0.046$ &            $5.312\pm0.030$ &         &    \nodata &    \nodata &    \nodata &    \nodata &      \nodata \\[2.0pt]
         &          $55\pm7$ &  $231^{+11}_{-12}$ &   $2203^{+97}_{-101}$ &    $62743^{+1950}_{-1992}$ &   00004 &    \nodata &    \nodata &    \nodata &    \nodata &      \nodata \\[5.5pt]
0950+139 &  $14.561\pm0.037$ &   $13.543\pm0.042$ &       $9.566\pm0.042$ &            $7.327\pm0.131$ &         &    \nodata &    \nodata &    \nodata &    \nodata &      \nodata \\[2.0pt]
         & $464^{+17}_{-18}$ &  $657^{+27}_{-28}$ &  $4724^{+193}_{-199}$ &     $9808^{+1124}_{-1266}$ &   11111 &  $930\pm3$ & $1151\pm3$ & $1703\pm8$ & $3726\pm9$ & $11853\pm45$ \\[5.5pt]
1821+643 &  $16.098\pm0.034$ &   $15.216\pm0.041$ &      $10.933\pm0.035$ &            $6.395\pm0.036$ &         &    \nodata &    \nodata &    \nodata &    \nodata &      \nodata \\[2.0pt]
         &         $113\pm4$ &          $141\pm6$ &    $1341^{+47}_{-48}$ &      $23140^{+830}_{-853}$ &   00000 &    \nodata &    \nodata &    \nodata &    \nodata &      \nodata \\[5.5pt]
\tablebreak
1958+015 &  $13.758\pm0.046$ &   $12.360\pm0.030$ &       $7.621\pm0.016$ &            $2.684\pm0.014$ &         &    \nodata &    \nodata &    \nodata &    \nodata &      \nodata \\[2.0pt]
         & $972^{+43}_{-45}$ & $1954^{+61}_{-62}$ & $28334^{+594}_{-598}$ & $705931^{+13897}_{-13974}$ &   00000 &    \nodata &    \nodata &    \nodata &    \nodata &      \nodata \\[5.5pt]
2333+301 &  $16.732\pm0.105$ &   $15.587\pm0.127$ &      $11.740\pm0.171$ &            $8.130\pm0.183$ &         &    \nodata &    \nodata &    \nodata &    \nodata &      \nodata \\[2.0pt]
         &          $63\pm6$ &  $100^{+11}_{-13}$ &    $638^{+93}_{-109}$ &       $4681^{+729}_{-862}$ &   00000 &    \nodata &    \nodata &    \nodata &    \nodata &      \nodata \\[5.5pt]
\enddata
\tablenotetext{a}{Spitzer Enhanced Imaging Products Source List (Cryogenic Release v2.0, January 2013) data types for the IRAC-1--4 and MIPS-1 bands, respectively, as follows:\ 0 = no data available, 1 = flux density measurement ($3.8^{\prime\prime}$ diameter aperture for IRAC, PSF-fit for MIPS), 2 = bandfill measurement (i.e., no source detection with S/N$>$3, so best combined position from detected bands is used to make a flux density measurement), 3 = 3-sigma upper limit (not used here), 4 = extended source, no photometry.}
\tablenotetext{b}{These values should be treated as upper limits -- see text for details.}
\tablecomments{Upper limits are $2\sigma$ (95\% confidence) levels.  The {\it Spitzer} flux density uncertainties do not include systematic error terms, which amount to an additional $\approx4.5$\% for IRAC and $\approx6.5$\% for MIPS.}
\end{deluxetable}

\clearpage
\begin{deluxetable}{cccccccccc}
\tabletypesize{\scriptsize}
\tablecolumns{9}
\tablewidth{0pt}
\tablecaption{Model Parameters for New WD Dust Disk Candidates \label{t:disks}}
\tablehead{
\colhead{WD} &
\colhead{Other Name} &
\colhead{$T_{\rm eff, wd}$} &
\colhead{distance} &
\colhead{$R_{\rm in, disk}$} &
\colhead{$R_{\rm out, disk}$} &
\colhead{inclination} &
\colhead{$\tilde{\chi}^{2}_{\rm wd}$\tablenotemark{a}} & 
\colhead{$\tilde{\chi}^{2}_{\rm all}$\tablenotemark{b}} & 
\colhead{$\tilde{\chi}^{2}_{\rm disk}$\tablenotemark{c}} 
\\
\colhead{ } &
\colhead{ } &
\colhead{(K)} &
\colhead{(pc)} &
\colhead{($R_{\rm wd}$)} &
\colhead{($R_{\rm wd}$)} &
\colhead{($^{\circ}$)} &
\colhead{ } & 
\colhead{ } & 
\colhead{ } 
}
\startdata
 0249-052 & HE~0245-0514  & 17823 & 104 & 30 & 52 & 80 &   1.3 &   5.1 &   1.9 \\
 0420-731 & \nodata       & 17653 &  79 & 18 & 88 & 71 &   2.8 & 151   &   6.1\tablenotemark{d} \\
 0420+520 & KPD~0420+5203 & 24301 &  76 & 12\tablenotemark{*} & 28 & 71 &   1.4 &  64   &   7.8\tablenotemark{d} \\
 0836+404 & DF~Lyn        & 11712 &  59 & 16 & 86 & 89 &   1.0 &   2.7 &   1.2 \\
 1046-017 & GD~124 & 14266 &  75 & 28 & 38 & 80 &  17.6\tablenotemark{e} &  18.0\tablenotemark{e} &  17.7\tablenotemark{e} \\
 1448+411 & CBS~204 & 13571 &  80 & 26 & 36 & 80 &   0.3 &   9.5 &   0.9\tablenotemark{d} \\
 2329+407 & EGGR~160 & 13900 &  33 & 28 & 68 & 80 & 107\tablenotemark{e}   & 155\tablenotemark{e}   &  94\tablenotemark{e} \\ 

          & &       &     &    &    &    &       &       &       \\
 
 1146-290 & Ruiz~440-146 &  5000 &  26 & \nodata & \nodata & \nodata &  12   &  32 & \nodata \\
 1330+473 & PG & 21223 &  91 & \nodata & \nodata & \nodata & 461   & 384 & \nodata \\
 2152-548 & 1ES~2152-54.8 & 45050 & 123 & \nodata & \nodata & \nodata &   0.7 &  10 & \nodata 
\enddata
\tablenotetext{*}{~Dust located at $R_{\rm in, disk}$ is at the assumed sublimation temperature.} 
\tablenotetext{a}{~$\tilde{\chi}^{2}$ value of the WD model fit compared to only the {\it GALEX} UV, optical, and 2MASS near-IR data points.}
\tablenotetext{b}{~$\tilde{\chi}^{2}$ value of the WD model fit compared to all available data points.}
\tablenotetext{c}{~$\tilde{\chi}^{2}$ value of the WD + dust disk model fit compared to all available data points.}
\tablenotetext{d}{~These values are large due to the inclusion of bright W3 and/or W4 points -- see discussion in the text.}
\tablenotetext{e}{~Large $\tilde{\chi}^{2}$ values due to poor fit in the UV -- see discussion in the text.}
\end{deluxetable}

\clearpage
\begin{deluxetable}{cccccccccc}
\tabletypesize{\scriptsize}
\setlength{\tabcolsep}{0.10in} 
\rotate
\tablecolumns{10}
\tablewidth{610pt}
\tablecaption{Published WD Dust Disk Model Parameters \label{t:pubdisks}}
\tablehead{
\colhead{WD} &
\colhead{Other Name} &
\colhead{Type\tablenotemark{a}} &
\colhead{$T_{\rm eff, wd}$} &
\colhead{$R_{\rm in, disk}$} &
\colhead{$R_{\rm out, disk}$} &
\colhead{inclination} &
\colhead{$T_{\rm subl}$} &
\colhead{Reference} &
\colhead{Notes} 
\\
\colhead{ } &
\colhead{ } &
\colhead{ } &
\colhead{(K)} &
\colhead{($R_{\rm wd}$)} &
\colhead{($R_{\rm wd}$)} &
\colhead{($^{\circ}$)} &
\colhead{(K)} &
\colhead{ } &
\colhead{ } 
}
\startdata

 0106-328   & HE 0106-3253                 &  DAZ & 15700 & 15 &      21 &      81 & \nodata    & \citet{fjl10} & 1 \\[3.0pt] % assuming Rwd = 0.013 Rsun 
 0110-565   & HE 0110-5630                 & DBAZ & 19200 & 31 &      35 &      60 & 1800       & \citet{gbf12} & \nodata \\[3.0pt]
 0146+187   & GD 16                        & DABZ & 11500 & 12\tablenotemark{*} &      30 &      48 & 1200       & \citet{fjz09} & \nodata \\   
            &                              &      & 11500 & 11 &      23 &      26 & 1000--1500 & \citet{jfz09} & 2 \\[3.0pt]  % Rout is transition radius from optically thick to thin
 0300-013   & GD 40                        &  DBZ & 15200 & 18\tablenotemark{*} &      44 &      78 & 1200       & \citet{jfz07a} & \nodata \\  
            &                              &      & 15200 & 13\tablenotemark{*} &      35 &      81 & 1000--1500 & \citet{jfz09} & 2 \\[3.0pt]  % Rout is transition radius from optically thick to thin
 0307+077   & HS 0307+0746                 &  DAZ & 10200 & 13 &      17 &      66 & \nodata    & \citet{fjl10} & 1 \\[3.0pt]  % assuming Rwd = 0.013 Rsun
 0408-041   & GD 56                        &  DAZ & 14400 & 16\tablenotemark{*} &     104 &       0 & 1200       & \citet{jfz07a} & \nodata \\   
            &                              &      & 14400 & 10\tablenotemark{*} &      71 &       0 & 1700       & \citet{jfz07a} & \nodata \\   
            &                              &      & 14400 & 16 &      45 &      45 & 2000       & \citet{vkk07} & 1 \\  % assuming Rwd = 0.013 Rsun
            &                              &      & 14400 & 30 &      65 &      41 & 1000--1500 & \citet{jfz09} & 2 \\[3.0pt]  % Rout is transition radius from optically thick to thin
 0435+410   & GD 61                        & DBAZ & 17280 & 19\tablenotemark{*} &      26 &      79 & 1300       & \citet{fbg11} & \nodata \\  
            &                              &      & 17500 & 17 &      36 &      85 & 1800       & \citet{gbf12} & \nodata \\[3.0pt]  
 J0738+1835 & SDSS J073842.56+183509.6     &  DBZ & 13600 &  9\tablenotemark{*} &      25 &      58 & 1800       & \citet{bgg12} & \nodata \\  
            &                              &      & 13600 & 12\tablenotemark{*} &      21 &       0 & 1400       & \citet{bgg12} & \nodata \\[3.0pt]  
 0842+231\tablenotemark{b} & Ton 345       &  DBZ & 18600 & 17\tablenotemark{*} &     100 &      66 & 1200--1500 & \citet{mja10} & \nodata \\  % gas+dust disk
            &                              &      & 18600 & 13\tablenotemark{*} &     187 &      83 & 1800       & \citet{bgg12} & \nodata \\[3.0pt]  
            &                              &      & 18600 & 19\tablenotemark{*} &     100 &      80 & 1400       & \citet{bgg12} & \nodata \\[3.0pt]  
 0843+516   & PG; SDSS J084702.28+512853.4 &   DA & 23900 & 30\tablenotemark{*} &      75 &      82 & 1200       & \citet{xj12} & \nodata \\[3.0pt]
 J0959-0200 & SDSS J095904.69-020047.6     &  DAZ & 13280 & 10\tablenotemark{*} &      25 &       0 & 1200       & \citet{fgs12} & \nodata \\[3.0pt]
 1015+161   & PG                           &  DAZ & 19300 & 24\tablenotemark{*} &      42 &      73 & 1200       & \citet{jfz07a} & \nodata \\[3.0pt]
 1041+091\tablenotemark{b} & SDSS J104341.53+085558.2     &  DAZ & 18330 & 23\tablenotemark{*} &      80 &      60 & 1200--1500 & \citet{mja10} & \nodata \\  % gas+dust disk
            &                              &      & 17912 & 13\tablenotemark{*} &      14 &      40 & 1800       & \citet{bgg12} & \nodata \\[3.0pt]  
            &                              &      & 17912 & 18\tablenotemark{*} &      38 &      85 & 1400       & \citet{bgg12} & \nodata \\[3.0pt]  
 1116+026   & GD 133                       &  DAZ & 12200 & 13\tablenotemark{*} &      83 &      78 & 1200       & \citet{jfz07a} & \nodata \\  
            &                              &      & 12200 & 12\tablenotemark{*} &      50 &      79 & 1000--1500 & \citet{jfz09} & 2 \\[3.0pt]  % Rout is transition radius from optically thick to thin
 1150-153   & EC11507-1519                 & DAVZ & 12800 & 10\tablenotemark{*} &      30 &       0 & 1000--1500 & \citet{jfz09} & 2 \\[3.0pt]  % Rout is transition radius from optically thick to thin
 J1221+1245 & SDSS J122150.81+124513.3     &  DAZ & 12250 & 11\tablenotemark{*} &      23 &      46 & 1200       & \citet{fgs12} & \nodata \\[3.0pt]
%
% 1225-079 = no published disk models?
%
 1226+110\tablenotemark{b} & SDSS J122859.92+104033.0     &  DAZ & 22000 & 18\tablenotemark{*} &     107 &      70 & 1200--1400 & \citet{bgm09} & \nodata \\  % gas+dust disk
            &                              &      & 22020 & 26 &      93 &      73 & 1200--1500 & \citet{mja10} & \nodata \\[3.0pt] 
 1349-230   & HE 1349-2305                 & DBAZ & 17000 & 13 &      35 &      85 & 1800       & \citet{gbf12} & \nodata \\[3.0pt]  
 1456+298   & G166-58                      &  DAZ &  7400 & 29 & \nodata & \nodata & 1200       & \citet{fzb08} & 3 \\[3.0pt]  % possible double degenerate binary with circumbinary disk
 1457-086   & PG                           &  DAZ & 20400 & 19\tablenotemark{*} &      21 &      73 & 1200       & \citet{fjz09} & \nodata \\[3.0pt]  
 1541+650   & KX Dra                       &  DAV & 11880 & 11 &      32 &      60 & \nodata    & \citet{kilic12} & \nodata \\[3.0pt]  
 J1557+0916 & SDSS J155720.77+091624.7     &  DAZ & 22810 & 25\tablenotemark{*} &      52 &      60 & 1200       & \citet{fgs12} & \nodata \\[3.0pt]
 J1617+1620 & SDSS J161717.04+162022.3     &   WD & 13432 &  9\tablenotemark{*} &      20 &      70 & 1800       & \citet{bgg12} & \nodata \\  
            &                              &      & 13432 & 12\tablenotemark{*} &      20 &      50 & 1400       & \citet{bgg12} & \nodata \\[3.0pt]  
 1729+371   & GD 362                       &  DAZ &  9740 & 12 & \nodata & \nodata & 1200       & \citet{jfz07b} & \nodata \\  
            &                              &      &  9740 &  6 &      38 &      60 & 2000       & \citet{vkk07} & \nodata \\[3.0pt]  
 1929+012   & GALEX J193156.8+011745       &  DAZ & 20890 & 23\tablenotemark{*} &      80 &      70 & 1350       & \citet{wired1} & \nodata \\  
            &                              &      & 23470 & 25\tablenotemark{*} &      40 & \nodata & 1400       & \citet{mfd11} & \nodata \\[3.0pt]  
 2115-560   & LTT 8452                     &  DAZ &  9700 & 13 &      25 &      80 & 2000       & \citet{vkk07} & \nodata \\  
            &                              &      &  9700 & 15 &      18 &      53 & 1200       & \citet{fjz09} & \nodata \\   
            &                              &      &  9700 &  8 &      30 &      74 & 1000--1500 & \citet{jfz09} & 2 \\[3.0pt]  % Rout is transition radius from optically thick to thin
 J2209+1223 & SDSS J220934.84+122336.5     &  DBZ & 17300 & 15\tablenotemark{*} &      45 &      57 & 1200       & \citet{xj12} & 4 \\  %  two degenerate disk models
            &                              &      & 17300 & 20\tablenotemark{*} &      60 &      40 & 1200       & \citet{xj12} & 4 \\[3.0pt]  % two degenerate disk models
 2221-165   & HE 2221-1630                 &  DAZ & 10100 & 11 &      21 &      60 & \nodata    & \citet{fjl10} & 1 \\[3.0pt]  % assuming Rwd = 0.013 Rsun
 2326+049   & G29-38                       &  DAZ & 11600 & 12 &      22 &      45 & 2000       & \citet{vkk07} & \nodata
\enddata
\tablecomments{
This table does not include the 52 WD dust disk candidates from \citet{wired2} -- see their Table~7.
(1) Published model radii given in units of R$_{\odot}$; we have assumed $R_{\rm wd}=0.013$~R$_{\odot}$.
(2) $R_{\rm out, disk}$ is transition radius from optically thick to thin.
(3) Possible double degenerate (WD+WD) binary with a circumbinary dust disk.
(4) These two disk models have degenerate parameters and produce comparably good fits to the data.
}
\tablenotetext{*}{~Dust located at $R_{\rm in, disk}$ is at the assumed sublimation temperature.} 
\tablenotetext{a}{~WD types from MSonline or SIMBAD.}
\tablenotetext{b}{~WD has a gas+dust disk.}
\end{deluxetable}

%%% END TABLE %%%%%%%%%%%%%%%%%%%%%%%%%%%%%%%%%

\end{document}